\documentclass[12pt]{article}
\usepackage[margin=2.5cm]{geometry}
\usepackage{amsmath}
\usepackage{amssymb}
\usepackage[utf8]{inputenc}
\usepackage{bbold}
\usepackage{xcolor}
\usepackage[makeroom]{cancel}
\usepackage{mathtools}
\usepackage[colorlinks, allcolors=cyan]{hyperref}

\usepackage[backend=biber, style=phys, sorting=none, citestyle=numeric-comp, url=false, eprint=false]{biblatex}
\appto{\bibsetup}{\sloppy}

\bibliography{bibliography}

\title{Unification of Conformal and Fuzzy Gravities with Internal Interactions resulting in SO(10) and a Possible Probe through Stochastic Gravitational Wave Background}
\begin{document}
\author{Gregory Patellis$^1$, Danai Roumelioti$^1$, Stelios Stefas$^1$, George Zoupanos$^{1,2,3,4}$}\date{}

\maketitle
\begin{center}
\itshape$^1$ Physics Department, National Technical University of Athens, Zografou Campus, 157 80, Zografou, Greece\\
\itshape$^2$ Max-Planck Institut f\'ur Physik, Boltzmannstr. 8, 85 748 Garching/Munich, Germany\\
\itshape$^3$ Universit\"at Hamburg, Luruper Chaussee 149, 22 761 Hamburg, Germany\\
\itshape$^4$  Deutsches Elektronen-Synchrotron DESY, Notkestra{\ss}e 85, 22 607, Hamburg, Germany
\end{center}

\begin{center}
\emph{E-mails: \href{mailto:grigorios.patellis@tecnico.ulisboa.pt}{grigorios.patellis@tecnico.ulisboa.pt}, \href{mailto:danai\_roumelioti@mail.ntua.gr}{danai\_roumelioti@mail.ntua.gr}, \href{mailto:dstefas@mail.ntua.gr}{dstefas@mail.ntua.gr}, \href{mailto:george.Zoupanos@cern.ch}{george.zoupanos@cern.ch}}
\end{center}

\begin{abstract}
The unification of conformal and fuzzy gravities with internal interactions is based  on the facts that i)  the tangent group of a curved manifold and the manifold itself do not necessarily have the same dimensions and ii) both gravitational theories considered here have been formulated in a gauge theoretic way. We review the gauge-theoretic approach of gravities, commenting in particular on their diffeomorphism invariance, and the construction of conformal and noncommutative (fuzzy) gravity using the gauge-theoretic framework. Based on an extension of the four-dimensional tangent group, unification of both gravities with the internal interactions is achieved. Both unified schemes are examined at 1-loop level considering suitable spontaneous symmetry breakings to a $SO(10)$ grand unified theory and consequently down to the Standard Model of particle physics through  four specific spontaneous symmetry breaking channels. Each channel is examined against proton lifetime experimental bounds and its observation potential through gravitational signal from cosmic strings production is discussed.
\end{abstract}

\section{Introduction}
\label{sec1}
The Unification of all Interactions was an ultimate expectation of many theoretical physicists, originating more than a century ago with the work of Kaluza and Klein \cite{Kaluza:1921, Klein:1926} by elaborating the notion of extra dimensions. An extension of the original ideas of Kaluza and Klein, by introducing multiple extra dimensions, produced a revival of the original scheme, when it was realised \cite{Kerner:1968, CHO1987358, Cho:1975sf}, that the non-abelian gauge theories could be necessary ingredients for the description of the Standard Model (SM) of Elementary Particle Physics. Specifically assuming that the description of the total space-time manifold takes the form of a direct product $M_D = M_4 \times B$, where $B$ is a compact Riemannian space with a non-abelian isometry group $S$, it was found that the dimensional reduction of the higher-dimensional gravity theory leads to gravity coupled to a Yang-Mills theory, based on $S$ as the gauge group and scalars in four dimensions. The most attractive feature of this scheme is the geometrical unification of gravity with the rest of the interactions and moreover, it provided a natural explanation of gauge symmetries. Unfortunately, this scheme also has problems, such as the fact that a classical ground state corresponding to the direct product structure of $M_D$ cannot be found. The most serious problem though, from the low-scale physics point of view, is that the inclusion of fermions in the original action does not lead to chiral fermions in four dimensions \cite{Witten:1983}. By adding Yang-Mills in the original action, the serious problems are resolved at the cost of losing the geometrical descriptions of all interactions. Introducing Yang-Mills fields in higher dimensions and considering that they are part of a Grand Unified Theory (GUT) together with a Dirac section \cite{Georgi1999, FRITZSCH1975193}, the restriction to obtain chiral fermions in four dimensions is limited to the requirement that the total dimension of space-time has to be of the form $4k + 2$ (see e.g., ref. \cite{CHAPLINE1982461}). At this point, it is worth noting that the schemes which will be mostly discussed in the present article go towards the other extreme, namely that all interactions can be described as gauge theories. Nevertheless, it should be emphasised that for several decades the Superstring theories (see e.g., refs. \cite{Green2012-ul, polchinski_1998, Lust:1989tj}) dominated the research involving the consideration of extra dimensions.

Superstring theories consist a solid framework, with the heterotic string theory \cite{GROSS1985253} which is defined in ten dimensions being the most attractive, in the sense that, in principle, they can at least be experimentally compatible, since the Standard Model gauge group can be accommodated into the gauge groups of the Grand Unified Theories that result after the dimensional reduction of the initial $E_8 \times E_8$ one. It is worth noting though, that even before the superstring theories era, another framework had been developed that focused on the dimensional reduction of higher-dimensional gauge theories, which has provided another direction for exploring the unification of fundamental interactions \cite{forgacs, MANTON1981502, Kubyshin:1989vd, KAPETANAKIS19924, LUST1985309}. The latter approach to unify fundamental interactions, which shared common objectives with the superstring theories, although less ambitious, was first examined in detail by Forgacs-Manton (F-M) and Scherk-Schwartz (S-S). F-M have established the concept of Coset Space Dimensional Reduction (CSDR) \cite{forgacs, MANTON1981502, Kubyshin:1989vd, KAPETANAKIS19924}, which can lead naturally to chiral fermions while S-S focused on the group manifold reduction \cite{SCHERK197961}, which although does not admit chiral fermions, its basic idea was used later as a prototype in many superstring model building attempts. Recent developments and attempts in the framework of CSDR towards realistic models that can be confronted with experiments can be found in refs. \cite{Manousselis_2004, Chatzistavrakidis:2009mh, Irges:2011de, Manolakos:2020cco, Patellis:2024dfl}. 

It should be added that in the attempts to achieve the goal of unification of all interactions, another direction has been developed directly in four dimensions based on the obvious interface of the notion of gauge invariance. Concretely, the Standard Model of Particle Physics is clearly based on gauge theories, while it was long known that gravity may be regarded as a gauge theory as well \cite{utiyama, kibble1961, Sciama, Umezawa, Matsumoto, macdowell, Ivanov:1980tw, Ivanov:1981wn, stellewest, Kibble:1985sn}. Therefore, it appeared as a very interesting challenge to further examine this relationship. One of the main reasons for the renewed interest in this subject before the superstrings period was the progress that was made in supergravity theories \cite{freedman_vanproeyen_2012, Ortín_2015}, which can very profitably be regarded as gauge theories. More recently, the interest was also directed in the noncommutative gravity \cite{castellani, Chatzistavrakidis_2018, Manolakos_paper1, manolakosphd, Manolakos_paper2, Manolakos:2022universe, Manolakos:2023hif, roumelioti2407}. 

Weyl \cite{weyl, weyl1929} was the first to associate electromagnetism with the phase transformations of the electron field, and moreover to develop the vierbein formalism, which appears to be very useful in gauge theories of gravity. A crucial step in the development of gravity as a gauge theory was done by Utiyama \cite{utiyama}, who demonstrated that gravity might be regarded as a gauge theory of the Lorentz group $SO(1,3)$, i.e. the spin connection could be treated as the gauge field of the theory, although the vierbein was introduced in a rather ad hoc way. The latter weakness was improved by Kibble \cite{kibble1961} and Sciama \cite{Sciama}, who considered instead the gauging of the Poincar\'e group. A further, more elegant development of the theory was done by Stelle and West \cite{stellewest, Kibble:1985sn} who identified the spin connection and the vierbein as parts of the gauge fields of the de Sitter (dS) $SO(1,4)$ or the Anti-de Sitter (AdS) group $SO(2,3)$, which was spontaneously broken by a scalar field to the Lorentz $SO(1,3)$ group. It is also worth noting the use of the gauge theory of the conformal group, $SO(2,4)$, in the construction of Weyl Gravity (WG) \cite{KAKU1977304, Roumelioti:2024lvn}, the Fuzzy Gravity (FG) \cite{Chatzistavrakidis_2018, Manolakos_paper1, manolakosphd, Manolakos_paper2, Manolakos:2022universe, Manolakos:2023hif, roumelioti2407} and its supersymmetric extension, the superconformal group, in the $N = 1$ supergravity \cite{KAKU1977304, freedman_vanproeyen_2012}.

Another interesting and more direct suggestion towards unifying gravity as a gauge theory with the other known interactions described by GUTs has been suggested in the past \cite{Weinberg:1984ke, Percacci:1984ai, Percacci_1991}, and revived recently \cite{Witten:1988hc, Nesti_2008, Nesti_2010, Chamseddine2010, Chamseddine2016, Krasnov:2017epi, Konitopoulos:2023wst, Manolakos:2023hif, noncomtomos, Roumelioti:2024lvn, Patellis:2024znm, Roumelioti:dubna, Roumelioti:2025cxi}. It is based on the observation that the tangent group of a curved manifold does not necessarily have the same dimension as the manifold. This possibility opens the very interesting avenue that one can consider higher-than-four-dimensional tangent groups in a four-dimensional spacetime and possibly achieve unification of gravity with internal interactions by gauging these higher-dimensional tangent groups. Then, to a great extent, the machinery that has been developed examining higher-dimensional theories with extra physical space dimensions, such as those used in the CSDR scheme \cite{CHAPLINE1982461, forgacs, MANTON1981502, Kubyshin:1989vd, KAPETANAKIS19924, LUST1985309, SCHERK197961, Manousselis_2004, Chatzistavrakidis:2009mh, Irges:2011de, Manolakos:2020cco, Patellis:2024dfl}, can be transferred in the four-dimensional constructions since they have the same tangent group. Examples of the latter are the constraints, such as the Weyl condition that has to be imposed in the higher-dimensional theories in order to result in realistic four-dimensional chiral theories describing the internal interactions after the dimensional reduction. Similarly, the need to impose the Majorana condition in addition to Weyl in the extra-dimensional theories in order to avoid a possible doubling of the spectrum of the reduced chiral theories \cite{CHAPLINE1982461, KAPETANAKIS19924}. Along this outlined direction, a unification of the gauge conformal group with those of the internal interactions has been constructed recently \cite{Manolakos:2023hif, Konitopoulos:2023wst, Patellis:2024znm, Roumelioti:2024lvn, Roumelioti:dubna, Roumelioti:2025cxi}. This construction was subsequently extended to the unification of the gauge conformal group on a covariant noncommutative (fuzzy) space with internal interactions \cite{roumelioti2407}.

In the present paper, we would firstly like to review the gauge theoretic approach of gravities, giving some emphasis on the conditions that guarantee the equivalence between gauge and diffeomorphism invariance. Then we review the construction of the Conformal Gravity and the Noncommutative (Fuzzy) Gravity using the gauge theoretic framework, while based on an extension of the four-dimensional tangent group we present the Unification of Conformal and Fuzzy Gravity with Internal Interactions. Finally, both unified schemes are further examined, concerning their behaviour at low energies, after suitable spontaneous symmetry breakings, and possible gravitational wave signals from the resulting cosmic string topological defects..

More specifically, in Section \ref{sec2} we present the resulting constraints from the requirement that the gauge invariance in the three-dimensional gauge gravity implies diffeomorphism invariance too. In Section \ref{sec3}, the four-dimensional Einstein Gravity (EG) as a gauge theory is presented. This is done by gauging the Anti-de Sitter group, $SO(2,3)$, which after spontaneous breaking leads to $SO(1,3)$, of the Einstein action with a cosmological constant. In Section \ref{sec4}, we discuss the gauging of the conformal group, $SO(2,4)$, and its breakings to Einstein and Weyl Gravities. In addition, we discuss how Weyl Gravity can break spontaneously to Einstein Gravity, and we comment on the conditions that guarantee the equivalence among gauge and diffeomorphism invariance. Section \ref{sec5} is devoted to the discussion of Noncommutative (Fuzzy) Gravity. Section \ref{sec6} deals with the Unification of Conformal and Fuzzy Gravities with Internal Interactions based on $SO(2,16)$. Finally, Section \ref{sec7} focuses on a 1-loop analysis of several breaking channels and discusses the $SO(10)$ breakings which satisfy proton lifetime bounds and can potentially provide gravitational signals.  Finally, in Section \ref{sec8} we summarise our results.

\section{Three-dimensional Einstein gravity as a Chern-Simons gauge theory}
\label{sec2}

In this section, we present the 3-d EG in order to establish that it can be described as a gauge theory, and more specifically as a Chern-Simons one \cite{Witten:1988hc}. The vielbein formalism and Palatini action are used in the appropriate 3-d version and therefore, the vielbein (assumed to be invertible) and the spin connection are treated as dynamical variables instead of the metric tensor.

In 3-d, for a manifold M, the Einstein-Hilbert (E-H) action in the vielbein formalism, without the inclusion of cosmological constant and matter, is: 
\begin{equation}
\label{3dEH}
     S_{EH_3}=\frac{1}{16\pi G}\int_M \epsilon^{
     \mu \nu \rho}e_{\mu}{}^{a} \left(\partial_{\nu} \omega_{\rho a} - \partial_{\rho} \omega_{\nu a} +\epsilon_{a b c} \omega_{\nu}{}^{ b}\omega_{\rho}{}^{c}\right).
 \end{equation}
Then, variation of the action with respect to $\omega$ provides us with the torsionless condition, i.e.
\begin{equation}
    T_{\mu \nu}^a=\partial_\mu e_\nu{}^a-\partial_\nu e_\mu{}^a+\epsilon^{a b c} \omega_{\mu b} e_{\nu c}-\epsilon^{a b c} \omega_{\nu b} e_{\mu c}=D_\mu e_\nu{}^a-D_\nu e_\mu{}^a=0
\end{equation}
where
\begin{equation}
\label{deriv3}
    D_\mu e_\nu{}^a=\partial_\mu e_\nu{}^a + \epsilon^{a b c} \omega_{\mu b} e_{\nu c}.
\end{equation}
In addition, variation with respect to $e$ gives the Einstein equations of motion in vacuum,
\begin{equation}
R_{\mu \nu a}=\partial_\nu \omega_{\rho}{}^{a}-\partial_\rho \omega_{\nu}{}^{a}+\epsilon_{a b c} \omega_\nu{}^b \omega_\rho{}^c=0\, .
\end{equation}
Note that in the 3-d case the redefinition $\omega_\mu{}^a = \frac{1}{2} \epsilon^{abc} \omega_{\mu bc}$ is permitted and is taken into account.

Collectively denoting the vielbein and spin connection as a gauge field $A$, the action can be written as $AdA + A^3$ , which is the general form of a Chern-Simons functional in 3-d. This in turn is suggestive towards a possible relation of the 3-d gravity with the Chern-Simons gauge theory in the same dimensions. It then suffices to find the appropriate gauge group and write down an action of Chern-Simons form and try to reach its coincidence with the 3-d E-H action, \eqref{3dEH}.

A first guess would be to consider the $ISO(1,2)$ to be the appropriate gauge group. Note though that the Chern-Simons functional is defined on simple Lie groups. Therefore, it is not straightforward to develop a Chern-Simons gauge theory of $ISO(1,2)$. What is required is to find an invariant quadratic form on the $ISO(1,2)$ Lie algebra. Indeed in 3-d for the $ISO(1,2)$ group there exists the following invariant and non-degenerate form,
\begin{equation}
tr(J_a P_b ) = \delta_{ab}\ ,\quad tr(P_a P_b ) = 0\ ,\quad tr(J_a J_b ) = 0\ ,
\end{equation}
where $J_a = \frac{1}{2} \epsilon_{abc} J^{bc}$ are the three Lorentz generators and $P_a$ are the three translations, which together accommodate the six generators of the $ISO(1,2)$ group. The above generators satisfy the following commutation relations:
\begin{equation}
\label{algebra3}
\left[J_a, J_b\right] = \epsilon_{abc} J^c\ ,\quad \left[J_a , P_b \right] = \epsilon_{abc} P^c\ ,\quad \left[P_a , P_b\right] = 0 .
\end{equation}
Then one can write down the gauge covariant derivative:
\begin{equation}
\label{covdev3}
    \tilde{D}_\mu=\partial_\mu+\left[A_\mu, \cdot \right],
\end{equation}
where $A_\mu(x)$ is the gauge connection expanded on the generators of $ISO(1,2)$,
\begin{equation}
\label{gaugeconnection3}
       A_\mu (x) = e_\mu{}^a(x) P_a + \omega_\mu{}^a(x) J_a\ ,
\end{equation}
i.e., a component gauge field has been assigned to each generator. The vielbein (dreibein) field has been corresponded to the local translations and the spin connection to the Lorentz transformations.

By construction $\tilde{D}_\mu$ transforms covariantly providing the transformation rule of $A_\mu$,
\begin{equation}
\label{gftransform3}
    \delta A_\mu=-\tilde{D}_{\mu}\epsilon=-\partial_\mu-\left[A_\mu,\epsilon\right]\ ,
\end{equation}
where $\epsilon = \epsilon(x)$ is the gauge transformation parameter, which, being an element of the $ISO(1,2)$ algebra, can be expanded on its generators,
\begin{equation}
\label{infparameter3}
    \epsilon(x)=\xi^a(x)P_a+\lambda^a(x) J_a\ ,
\end{equation}
with $\xi^a(x)$ and $\lambda^a(x)$ being infinitesimal parameters. Then from eqs \eqref{covdev3}, \eqref{gftransform3} and \eqref{infparameter3} and using the algebra of generators \eqref{algebra3} one finds the transformations of the fields $e$ and $\omega$,
\begin{align}
\label{deltae}
\delta e_\mu{}^a & =-\partial_\mu \xi^a-\epsilon^{a b c} e_{\mu b} \lambda_c-\epsilon^{a b c} \omega_{\mu b} \xi_c\, , \\
\label{deltaomega}
\delta \omega_\mu{}^a & =-\partial_\mu \lambda^a-\epsilon^{a b c} \omega_{\mu b} \lambda_c \ .
\end{align}

The next step in constructing of the action of the gauge theory of $ISO(1,2)$ is to determine the tensors of the gauge fields using the commutator of the covariant derivative of the gauge theory, $\tilde{D}_\mu$, i.e.
\begin{equation}
\label{Rdef3}
R_{\mu \nu}=\left[\tilde{D}_\mu, \tilde{D}_\nu\right]=\partial_\mu A_\nu-\partial_\nu A_\mu+\left[A_\mu, A_\nu\right]\ ,
\end{equation}
with $A_\mu$ the gauge connection \eqref{gaugeconnection3}. Taking into account that $R_{\mu\nu}$ is valued in the algebra of $ISO(1,2)$, we can also write its expansion on the generators of the algebra,
\begin{equation}
\label{Rxpansion3}
R_{\mu \nu}=T_{\mu \nu}{}^a(x) P_a+R_{\mu \nu}{}^a(x) J_a\ .
\end{equation}
From eqs \eqref{gaugeconnection3}, \eqref{Rdef3} and \eqref{Rxpansion3}, one obtains the expressions of the component curvature tensors,
\begin{align}
\label{torsion3}
& T_{\mu \nu}{}^a=\partial_\mu e_\nu{}^a-\partial_\nu e_\mu{}^a+\epsilon^{a b c} \omega_{\mu b} e_{\nu c}-\epsilon^{a b c} \omega_{\nu b} e_{\mu c}\, , \\
\label{curvature3}
& R_{\mu \nu}{}^a=\partial_\mu \omega_{\nu a}-\partial_\nu \omega_{\mu a}+\epsilon_{a b c} \omega_\mu{}^b \omega_\nu{}^c\, ,
\end{align}
which are the 3-d versions of the torsion and curvature two-forms.

As is already mentioned, the gauge theory in 3-d to be constructed is the Chern-Simons action functional,
\begin{equation}
S_{\mathrm{CS}}=\int_M \operatorname{tr}(A \wedge d A+A \wedge A \wedge A)=\int_M \operatorname{tr} A_\mu\left(\partial_\nu A_\rho-\partial_\rho A_\nu+\left[A_\nu, A_\rho\right]\right) \epsilon^{\mu \nu \rho} d^3 x\ ,
\end{equation}
which, upon substituting the expression for $A_\mu$ from \eqref{gaugeconnection3}, becomes:
\begin{equation}
\int_M \epsilon^{\mu \nu \rho} e_{\mu}{}^{a} \left((\partial_{\mu} \omega_{\rho a} - \partial_{\rho} \omega_{\nu a} + \omega_{\nu}{}^{b} \omega_{\rho}{}^{c} \epsilon_{abc}) + (\partial_{\nu} e_{\rho a} - \partial_{\rho} e_{\nu a} + (\omega_{\nu}{}^{b} e_{\rho}{}^{c} - e_{\nu}{}^{b} \omega_{\rho}{}^{c}) \epsilon_{abc}) 
\right)\ ,
\end{equation}
i.e. the action is expressed in terms of the torsion and curvature two-forms given in eqs \eqref{torsion3} and \eqref{curvature3} respectively. Aiming for Lorentz, $SO(1,2)$ invariance of the action, the torsionless condition has to be imposed and therefore the final action takes the form,
\begin{equation}
\label{3dCSaction}
S_{\text{CS}} = \int_M \epsilon^{\mu \nu \rho} e_{\mu}{}^{a} 
\left( \partial_{\nu} \omega_{\rho a} - \partial_{\rho} \omega_{\nu a} + \omega_{\nu}{}^{b} \omega_{\rho}{}^{c} \epsilon_{abc} \right)\ ,
\end{equation}
which, up to a constant, coincides with the E-H action in 3-d.

Clearly, by construction, the action \eqref{3dCSaction} is invariant under the gauge transformations of the component fields $e$ and $\omega$, as they are given in eqs \eqref{deltae} and \eqref{deltaomega}. The important point we would like to stress here is that one can show that the gauge transformations are equivalent to the diffeomorphism transformations. In other words, the gauge transformations of the fields compensate for the coordinate transformations in the present gauge-theoretic approach.

In order to prove the above statement, let us consider the transformations of the vielbein and the spin connection under a diffeomorphism generated by a vector field, $v^\nu$. The parametrization of these transformations, denoted by $\tilde{\delta}e_\mu{}^a$ and $\tilde{\delta}\omega_\mu{}^a$ , are given by the Lie derivatives in the direction of the vector field $v^\nu$,
\begin{align}
    \tilde{\delta} e_{\mu}{}^{a} &= \mathcal{L}_{-v} e_{\mu}{}^{a} = -v^{\nu} \partial_{\nu} e_{\mu}{}^{a} - (\partial_{\mu} v^{\nu}) e_{\nu}{}^{a}= -v^{\nu} (\partial_{\nu} e_{\mu}{}^{a} - \partial_{\mu} e_{\nu}{}^{a}) - \partial_{\mu} (v^{\nu} e_{\nu}{}^{a}) ,\label{eGCT} \\
    \tilde{\delta} \omega_{\mu}{}^{a} &= \mathcal{L}_{-v} \omega_{\mu}{}^{a} = -v^{\nu} \partial_{\nu} \omega_{\mu}{}^{a} - (\partial_{\mu} v^{\nu}) \omega_{\nu}{}^{a}= -v^{\nu} (\partial_{\nu} \omega_{\mu}{}^{a} - \partial_{\mu} \omega_{\nu}{}^{a}) - \partial_{\mu} (v^{\nu} \omega_{\nu}{}^{a}) .
\end{align}

Now, we can write down the difference between the above diffeomorphism and the gauge transformation of $e_\mu{}^a$, $\tilde{\delta}e_\mu{}^a -\delta e_\mu{}^a$, introducing $\xi^a=v^\nu e_\nu{}^a $ and $\lambda^a= v^\nu\omega_\nu{}^a$
\begin{equation}
\label{deltaE3d}
\begin{aligned}
    \tilde{\delta}e_\mu{}^a -\delta e_\mu{}^a &=-v^\nu(\partial_\nu e_\mu{}^a-\partial_\mu e_\nu{}^a)- \partial_\mu(v^\nu e_\nu{}^a)+\partial_\mu(e_\nu{}^a v^\nu)+\epsilon^{abc}e_{\mu b}v^\nu\omega_{\nu c}+\epsilon^{abc}\omega_{\mu b}v^\nu e_{\nu c}\\
    &=-v^\nu(D_\nu e_\mu{}^a-D_\mu e_\nu{}^a)\ ,
\end{aligned}
\end{equation}
where the definition of $D_\mu$ is given in \eqref{deriv3}. It is now evident that the above difference vanishes given the constraint of the torsionless condition, which was imposed in the action. Correspondingly the difference $\tilde{\delta}\omega_\mu{}^a -\delta \omega_\mu{}^a$, with $\lambda^a =v^\nu \omega_\nu{}^a $ as before, is found to be
\begin{equation}
\label{deltaOmega3d}
\begin{aligned}
    \tilde{\delta}\omega_\mu{}^a -\delta \omega_\mu{}^a &=-v^\nu(\partial_\nu \omega_\mu{}^a-\partial_\mu \omega_\nu{}^a)- \partial_\mu(v^\nu \omega_\nu{}^a)+\partial_\mu(v^\nu \omega_\nu{}^a )+\epsilon^{abc}\omega_{\mu b}v^\nu\omega_{\nu c}\\
    &=v^\nu(\partial_\mu \omega_\nu{}^a-\partial_\nu \omega_\mu{}^a+\epsilon^{abc}\omega_{\mu b}\omega_{\nu c})=v^\nu R_{\mu \nu}\ ,
\end{aligned}
\end{equation}
which in turn vanishes by the equation of motion. Therefore, it is shown that the gauge transformations are equivalent to the diffeomorphism transformations on-shell and as a result the gauge transformations of the fields compensate those of the coordinate transformations. Stating the result more clearly, the invariance of the action under the gauge transformations ensures the general covariance of the theory. In addition, it is confirmed that the appropriate group for constructing the 3-d gravity as a gauge theory is the $ISO(1,2)$.

\section{Four-dimensional Einstein gravity as a gauge theory}
\label{sec3}

The way the 4-d gravity is described as a gauge theory is less straightforward than in 3-d, discussed in Section \ref{sec2}. Denoting again collectively the vielbein and spin connection as a gauge field A, as it was done in the 3-d case, the expression of the E-H action would have the form $\sim A \wedge A \wedge (dA + A^2)$ and such an action cannot be reproduced by a gauge theory. However, one could build the Ricci scalar invariant using the curvature tensor, $R_{\mu \nu}{}^{ab}$ and an action including this invariant could be constructed recovering the E-H action \cite{Ramond:1981pw}. However, there is still another indirect way to result with E-H action treating the Lorentz and translational part in a unified way, based on intuitive and physical arguments. The idea is to start from an action with larger gauge symmetry than the Lorentz and  break it to $SO(1,3)$ either by constraints or spontaneously (SSB) employing a scalar field. The natural choice of the larger gauge group to be spontaneously broken to $SO(1,3)$ would be the Poincar\'e group, $ISO(1, 3)$. However, the distinct behaviour of the translation generators does not allow a full local gauge symmetry, as is the case of Lorentz transformations, thus another group has to be employed, in which all the generators would be on equal footing. The two candidate groups that fulfill this additional requirement are the dS, $SO(1,4)$, and the AdS, $SO(2,3)$, groups, which moreover, both contain the same number of generators as the Poincar\'e group. Such actions have already been constructed, \cite{stellewest, Kibble:1985sn, manolakosphd, Roumelioti:2024lvn} and following the SSB, they lead to the E-H action with cosmological constant.

Let us be more explicit on the above. First, as in the 3-d case described in Section \ref{sec2}, the vierbein formalism has to be employed for the construction of the gauge theory of gravity. In absence of cosmological constant, the isometry group (symmetries of the metric) of the Minkowski space-time is $ISO(1,3)$ (the Poincar\'e group) and it is the one that will be considered as the gauge group, in accordance with the three-dimensional case, where isometry groups of the Minkowski, dS and AdS spaces were considered as the gauge groups. The Poincar\'e algebra comprises ten generators, four local translations, $P_a$ and six Lorentz transformations, $M_{ab}$ , satisfying the following commutation relations\footnote{The bracket notation implies the antisymmetricity of the indices inside, i.e. $\eta_{a[b}P_{c]}=\frac{1}{2}(\eta_{ab}P_{c}-\eta_{ac}P_{b})$.}:
\begin{equation}
[M_{ab},M_{cd}]=4\eta_{[a[c}M_{d]b]},\quad[P_{a},M_{bc}]=2\eta_{a[b}P_{c]},\quad[P_{a},P_{b}]=0,
\end{equation}
where $\eta_{ab} = diag(-1, +1, +1, +1)$ is the four-dimensional Minkowski metric. Following the standard procedure, the gauge covariant derivative is defined as:
\begin{equation}
D_\mu=\partial_\mu+[A_\mu,\cdot],
\end{equation}
where $A_\mu(x)$ is the gauge connection. Expansion of the connection on the generators of $ISO(1,3)$ gives the expression:
\begin{equation}
\label{2.65}
    A_\mu(x)=e_\mu{}^a(x)P_a+\frac{1}{2}\omega_\mu{}^{ab}(x)M_{ab}\,,
\end{equation}
where $e_\mu{}^a$ and $\omega_\mu{}^{ab}$ are identified as the component gauge fields for the translations and Lorentz transformations respectively. By definition, the transformation of $D_\mu$ is covariant, therefore, the transformation law for the gauge connection $A_\mu$ is given by:
\begin{equation}
\label{2.66}
    \delta A_\mu=D_\mu\epsilon=\partial_\mu+[A_\mu,\epsilon]\,,
\end{equation}
where $\epsilon=\epsilon(x)$ is a gauge transformation parameter, which, as an element of $ISO(1,3)$ algebra, it may be written as an expansion on the generators:
\begin{equation}
\label{2.67}
    \epsilon(x)=\xi^a(x)P_a+\frac{1}{2}\lambda^{ab}(x)M_{ab}\,,
\end{equation}
with $\xi^a(x)$ and $\lambda^{ab}(x)$ being infinitesimal parameters. Combination of \eqref{2.65}, \eqref{2.66} and \eqref{2.67} leads to the expression of the transformation of the component gauge fields:
\begin{align}
    \delta e_{\mu}{}^{a}&=\partial_{\mu}\xi^{a}+\omega_{\mu}{}^{ab}\xi_{b}-\lambda^a{}_{b}e_{\mu}{}^{b}\,,\\
    \delta\omega_{\mu}{}^{ab}&=\partial_{\mu}\lambda^{ab}+\lambda^{a}{}_c\omega_{\mu}{}^{bc}-\lambda^{b}{}_{c}\omega_{\mu}{}^{ac}\,.
\end{align}
The corresponding field strength tensors, $T_{\mu\nu}{}^a$ and $R_{\mu\nu}{}^{ab}$, of the component fields, $e$ and $\omega$, are obtained by the definition of the field strength tensor, $R_{\mu\nu}$, of $A_\mu$:
\begin{equation}
\label{2.70}
    R_{\mu\nu} = [D_\mu, D_\nu] = \partial_\mu A_\nu - \partial_\nu A_\mu + [A_\mu, A_\nu]\, ,
\end{equation}
after its expansion on the generators:
\begin{equation}
\label{2.71}
    R_{\mu\nu} = T_{\mu\nu}{}^a P_a + \frac{1}{2} R_{\mu\nu}{}^{ab} M_{ab}\, .
\end{equation}
Therefore, combining \eqref{2.65}, \eqref{2.70} and \eqref{2.71}, the expressions of the component tensors are:
\begin{align}
    T_{\mu\nu}{}^a &= \partial_\mu e_\nu{}^a - \partial_\nu e_\mu{}^a - \omega_\mu{}^{ab} e_{\nu b} + \omega_\nu{}^{ab} e_{\mu b}\, ,\\
    R_{\mu\nu}{}^{ab} &= \partial_\mu \omega_\nu{}^{ab} - \partial_\nu \omega_\mu{}^{ab} - \omega_\mu{}^{ac} \omega_\nu{}^{b}{}_{c} + \omega_\nu{}^{ac} \omega_\mu{}^b{}_c\, ,
\end{align}
where the above expressions coincide with the ones found for the torsion and curvature two-forms in the vierbein formalism description of general relativity.

So far, the construction of the gauge-theoretic version of 4-d gravity is done in a straightforward way. Going further to the dynamical part of the theory, the obvious choice would be an action of Yang-Mills type of the Poincaré group. However, in order to claim a successful relation between the four-dimensional gravity and a gauge theory, it is necessary to result with the Einstein-Hilbert action, which is, of course, not of the usual Yang-Mills type, as already mentioned. To proceed, let us note that the desired action has to be invariant under the Lorentz transformations and not under the total Poincar\'e symmetry. Moreover, the distinct behaviour of the translations generators does not allow a full local gauge symmetry, as is the case of Lorentz transformations, thus another group is needed to be employed, in which all the generators would be on equal footing. The two candidate groups, as already mentioned, are the dS, $SO(1, 4)$, and the AdS group, $SO(2, 3)$. Both of these groups can be spontaneously broken to the Lorentz group, $SO(1, 3)$, and both contain the same number of generators as the Poincar\'e group, but with the property that they can be denoted by a single gauge field, $\omega_\mu{}^{AB}$, where $A, B = 1,\dots,5$. Here, following the approach discussed in \cite{stellewest, kibble1961, Roumelioti:2024lvn}, we are going to summarise the case of $SO(2, 3)$, the AdS group. The case of $SO(1, 4)$ has been presented in \cite{manolakosphd}. The algebra of the $SO(2, 3)$ group is the following:
\begin{equation}
    \left[J_{A B}, J_{C D}\right] = \eta_{BC}J_{AD}+\eta_{AD} J_{BC} -\eta_{AC} J_{BD} -\eta_{BD} J_{AC} ,
\end{equation}
where the metric of the gauge theory is $\eta_{AB}=\operatorname{diag}(-1,1,1,1,-1)$ with $A,B=1,\dots,5$, while the gauge connection for the gauge fields, $\omega_\mu{}^{AB}$, is $A_\mu = \frac{1}{2}\omega_\mu{}^{AB}J_{AB}$, where $J_{AB}$ are the ten generators of the AdS group. Employing the definition of the field strength tensor of $A_\mu$, 
\begin{equation}
F_{\mu \nu}=\left[D_\mu, D_\nu\right]=\partial_\mu A_\nu-\partial_\nu A_\mu+\left[A_\mu, A_\nu\right],
\end{equation}
and since $F_{\mu\nu} = \frac{1}{2}F_{\mu\nu}{}^{AB}J_{AB}$, we result with the expression
\begin{equation}
F_{\mu \nu}{}^{A B}=\partial_\mu \omega_\nu{}^{A B}-\partial_\nu \omega_\mu{}^{A B}+\omega_\mu{}^A{}_C \omega_\nu{}^{C B}-\omega_\nu{}^A{}_C \omega_\mu{}^{C B}.
\end{equation}
The only invariant that can be constructed that is also polynomial to the field strength tensor is the topological invariant that yields the Pontryagin index, 
\begin{equation}\label{initialaction}
S\sim \int d^4 x \epsilon^{\mu \nu \rho \sigma} F_{\mu \nu}{}^{A B} F_{\rho \sigma A B},
\end{equation} 
where $\epsilon^{\mu \nu \rho \sigma}$ is the Levi-Civita symbol. The above integral is a total divergence and also parity violating and C conserving (hence CP violating). Although it has been shown that there exists a non-polynomial choice of Lagrangian \cite{West:1978nd}, in the works of \cite{stellewest} and \cite{Kibble:1985sn} this choice was not eventually promoted due to the impractical expression of the integral. Instead, another choice is preferred, which is also of second order in terms of the field strength tensor, but in this case in addition to the gauge field an auxiliary scalar field, $y^A$, is introduced, along with a dimensionful  parameter, $m$, satisfying a constraint equation, while contractions are being done by using the tensor $\epsilon_{ABCDE}$. In this case the action is parity conserving, it is not a total divergence, as will be shown below, and it takes the form:
\begin{equation}
S=a_{AdS}\int d^4 x\left(m y^E \epsilon_{A B C D E} \frac{1}{4}F_{\mu \nu}{}^{A B} F_{\rho \sigma}{}^{C D}\epsilon^{\mu \nu \rho \sigma} +\lambda\left(y^E y_E+m^{-2}\right)\right),
\end{equation}
where $a_{AdS}$ is a dimensionless coupling, and $\lambda$ is a parameter acting as a Lagrange multiplier, which imposes the constraint
\begin{equation}\label{gauge}
    y^Ey_E=-m^{-2}.
\end{equation}
Picking a specific gauge in which,
\begin{equation}
    y=y^0=(0,0,0,0,m^{-1}),
\end{equation}
the $SO(2,3)$ symmetry is spontaneously broken down to the little
group (isotropy subgroup) of $y^0$, i.e. the Lorentz group, $SO(1,3)$\footnote{The reader is reminded that in the initial $SO(2,3)$ metric, $\eta_{AB}$, the 1st and the 5th component where chosen to be the timelike components.}. Since $y^0$ is timelike, then $m^2>0$.

The action of the remaining symmetry, then becomes 
\begin{equation}
S_{\mathrm{SO}(1,3)}=\frac{a_{AdS}}{4}\int d^4 x \epsilon^{\mu \nu \rho \sigma} F_{\mu \nu}{}^{a b} F_{\rho \sigma}{}^{c d} \epsilon_{a b c d}.
\end{equation}
Defining the scaled gauge field, $e_\mu{}^a=m^{-1} \omega_\mu{}^{a 5}$, which corresponds to the five broken translation-like generators, from the components of the initial symmetry field strength tensor we obtain:
\begin{equation}\label{tors}
F_{\mu \nu}{}^{a 5}=m (\partial_\mu e_\nu{}^a-\partial_\nu e_\mu{}^a-\omega_\mu{}^{a b} e_{\nu b}+\omega_\nu{}^{a b} e_{\mu b}),
\end{equation}
and
\begin{equation}\label{curv}
F_{\mu \nu}{}^{a b}=(\partial_\mu \omega_\nu{}^{a b}-\partial_\nu \omega_\mu{}^{a b}+\omega_\mu{}^{a}{}_{c} \omega_{\nu}{}^{cb}-\omega_\nu{}^{a}{}_{c} \omega_{\mu}{}^{cb})+m^2\left(e_\mu{}^a e_\nu{}^b-e_\nu{}^a e_\mu{}^b\right).
\end{equation}
The expression in parentheses in eq. \eqref{tors} is identified with the torsion tensor, $T_{\mu \nu}{}^a$, in the vielbein formalism description of GR, while the expression in the first parentheses of eq. \eqref{curv} is the curvature two-form, $R_{\mu \nu}{}^{ab}$. The absence of a $F_{\mu \nu}{}^{a 5}$ term in the broken action, implies that the theory is torsion-free, since it can be set equal to zero. Substituting the curvature two-form in the broken action, one obtains
\begin{equation}
    \begin{aligned}
        S_{\mathrm{SO}(1,3)} &=\frac{a_{AdS}}{4}\int d^4 x \epsilon^{\mu \nu \rho \sigma} \epsilon_{a b c d}\left[R_{\mu \nu}{}^{a b}+m^2\left(e_\mu{}^a e_\nu{}^b-e_\mu{}^b e_\nu{}^a\right)\right]\\
        &\quad\qquad\qquad\qquad\times\left[R_{\rho \sigma}{}^{c d}+m^2\left(e_\rho{}^c e_\sigma{}^d-e_\rho{}^d e_\sigma{}^c\right)\right] \\
        &=\frac{a_{AdS}}{4}\int d^4 x \epsilon^{\mu \nu \rho \sigma} \epsilon_{a b c d}\left[R_{\mu \nu}{}^{a b}R_{\rho \sigma}{}^{c d}+4m^2 R_{\mu \nu}{}^{a b}e_\rho{}^c e_\sigma{}^d+4m^4e_\mu{}^a e_\nu{}^b e_\rho{}^c e_\sigma{}^d\right].
    \end{aligned}
\end{equation}

From the above expression it is obvious that the resulting action consists of three terms of the general form
\begin{equation}
    S_{\mathrm{SO}(1,3)}=\frac{a_{AdS}}{4}\int d^4 x \epsilon^{\mu \nu \rho \sigma} \epsilon_{a b c d}\left(\mathcal{L}_{R R}+ m^2 \mathcal{L}_{R e e}+ m^4\mathcal{L}_{eeee}\right),
\end{equation}
where $a_{AdS}>0$. The first term does not contribute to the field equations, as it yields the Gauss-Bonnet (G-B) topological invariant. The second term is the one identified with the E-H action, as it contains the Ricci scalar curvature, while the third term is a cosmological constant of order $m^4$, which causes the maximally symmetric solution of the field equations to be the AdS space\footnote{Recall that the expression of the E-H Lagrangian in the presence of a cosmological constant is $\mathcal{L}_{\text{E-H}}\sim R-2\Lambda$, thus in our case the cosmological constant term is negative, as expected for the AdS space.},
\begin{equation}
F_{\mu \nu}{}^{a b}=0 \Rightarrow R_{\mu \nu}{}^{a b}=- m^2\left(e_\mu{}^a e_\nu{}^b-e_\nu{}^a e_\mu{}^b\right).
\end{equation}

Concluding, although the transformations of the gauge fields $e$ and $\omega$ can be obtained when choosing the Poincar\'e group as the gauge group, in order to result with an E-H equivalent action, one has to employ either the AdS or the dS group, starting from an action that is polynomial with respect to the field strength tensor, and include an auxiliary scalar field which satisfies a constraint and can be chosen to lie in a fixed gauge. Thus, a SSB is being induced and the original symmetry is being reduced to the Lorentz one. The resulting action, although gauge invariant, is not a total divergence, contrary to \eqref{initialaction}, and despite the fact that it includes the G-B topological term, the rest of its terms can provide the usual equations of motion of EG in AdS space. In this way, we manage to describe the four-dimensional EG as a gauge theory.

Concerning general covariance, it is recovered by the relation between gauge transformations and diffeomorphisms. Following the same procedure and calculations as in the 3-d case, one ends up with the four-dimensional versions of \eqref{deltaE3d} and \eqref{deltaOmega3d}. Therefore, taking into account the torsionless condition and the equation of motion of vanishing curvature, general covariance is ensured.

\section{Four-dimensional Conformal gravity}
\label{sec4}

The gauge-theoretic procedure has also been applied in the case of the conformal group, $SO(2,4)$ \cite{KAKU1977304, Fradkin:1985am, Maldacena:2011mk, mannheim, Anastasiou:2016jix, ghilencea2023, Hell:2023rbf, Condeescu:2023izl}. In the early constructions of the conformal group action, the authors imposed constraints to obtain WG. In the current work, as in our previous work \cite{Roumelioti:2024lvn}, we instead use the SSB mechanism. This can be done either by introducing two scalars in the vector representation of $SO(6) \sim SO(2,4)$ or by a scalar in the 2nd rank antisymmetric rep, $\textbf{15}$. In both cases, one can obtain either the $SO(1,3)$ gauge group or the WG.

The gauge group, $SO(2, 4)$ contains fifteen generators: six Lorentz transformations, $M_{ab}$, four translations, $P_{a}$ , four special conformal transformations (conformal boosts), $K_{a}$ and the dilatation (scale transformation), $D$. These generators satisfy the following commutation relations which determine the $SO(2,4)$ algebra,
\begin{equation}\label{24}
\begin{aligned}
[M_{ab},M_{cd}] &=\eta_{bc}M_{ad}+\eta_{ad}M_{bc}-\eta_{ac}M_{bd}-\eta_{bd}M_{ac}, \\
[M_{ab},P_{c}] &=\eta_{bc}P_{a}-\eta_{ac}P_{b}, \\
[M_{ab},K_{c}] &=\eta_{bc}K_{a}-\eta_{ac}K_{b}, \\
[P_{a},D] &=P_{a}, \\
[K_{a},D] &=-K_{a}, \\
[K_a,P_b] &=-2(\eta_{ab}D+M_{ab}),
\end{aligned}
\end{equation}
where $\eta_{ab}$ is the mostly positive four-dimensional Minkowski metric. The gauge connection, $A_\mu$, as an element of the $SO(2,4)$ algebra, can be expanded in terms of the generators as
\begin{equation}\label{25}
A_{\mu}=e_{\mu}{}^{a}P_{a}+\frac{1}{2}\omega_{\mu}{}^{ab}M_{ab}+b_{\mu}D+f_{\mu}{}^{a}K_{a},
\end{equation}
where, for each generator a gauge field has been introduced. The gauge fields related to the translations are identified as the vierbein, while those of the Lorentz transformations are identified as the spin connection. The gauge connection, $A_\mu$ , obeys the following infinitesimal transformation rule,
\begin{equation}\label{26}
\delta A_\mu=D_\mu\epsilon=\partial_\mu\epsilon+[A_\mu,\epsilon],
\end{equation}
where $\epsilon = \epsilon(x)$ is a parameter that belongs to the gauge algebra and for this reason it can be expanded too as,
\begin{equation}
\epsilon=\xi^{a}P_{a}+\frac{1}{2}\lambda^{ab}M_{ab}+\kappa D+\rho^{a}K_{a}.
\end{equation}
Combining the relations \eqref{24}, \eqref{25} and \eqref{26} one can obtain the transformation properties of the various gauge fields,
\begin{align}
\delta e_{\mu}{}^{a} & =\partial_{\mu}\xi^{a}+\omega_{\mu}{}^{a}{}_{b}\xi^{b}-b_{\mu}\xi^{a}-\lambda^{a}{}_{b}e_{\mu}{}^{b}+\kappa e_{\mu}{}^{a}, \label{deltaee} \\
\delta\omega_{\mu}{}^{ab} & =\partial_{\mu}\lambda^{ab}-2\omega_{\mu}{}^{ac}\lambda^b{}_{c}-4f_{\mu}{}^{[a}\xi^{b]}-4e_{\mu}{}^{[a}\rho^{b]}, \\
 \delta b_{\mu} & =\partial_{\mu}\kappa-2\xi^{a}f_{\mu a}+2\rho^{a}e_{\mu a}, \\
\delta f_{\mu}^{a} & =\partial_{\mu}\rho^{a}+\omega_{\mu}{}^{ab}\rho_{b}+b_{\mu}\rho^{a}-\lambda^{ab}f_{\mu b}-\kappa f_{\mu}{}^{a}.
\end{align}
The field strength tensor of the theory can be found in the standard way to be,
\begin{equation}
\label{fst}
F_{\mu\nu}=\tilde{R}_{\mu\nu}{}^{a}P_{a}+\frac{1}{2}R_{\mu\nu}{}^{ab}M_{ab}+R_{\mu\nu}D+R_{\mu\nu}{}^{a}K_{a}\, ,
\end{equation}
where its components are given as follows,
\begin{align}
R_{\mu\nu}{}^{ab}&=\partial_{\mu}\omega_{\nu}{}^{ab}-\partial_{\nu}\omega_{\mu}{}^{ab}-\omega_{\mu}{}^{ac}\omega_{\nu c}{}^{b}+\omega_{\nu}{}^{ac}\omega_{\mu c}{}^{b}-8e_{[\mu}{}^{[a}f_{\nu]}{}^{b]} \nonumber \\
    & =R_{\mu\nu}^{(0)ab}-8e_{[\mu}{}^{[a}f_{\nu]}{}^{b]}, \label{Rmnab} \\
\tilde{R}_{\mu\nu}{}^{a}&=\partial_{\mu}e_{\nu}{}^{a}-\partial_{\nu}e_{\mu}{}^{a}+\omega_{\mu}^{ab}e_{\nu b}-\omega_{\nu}{}^{ab}e_{\mu b}-2b_{[\mu}e_{\nu]}{}^{a}\nonumber \\
    & =T_{\mu\nu}^{(0)a}(e)-2b_{[\mu}e_{\nu]}{}^{a}, \\
R_{\mu\nu}{}^{a}&=\partial_{\mu}f_{\nu}{}^{a}-\partial_{\nu}f_{\mu}{}^{a}+\omega_{\mu}{}^{ab}f_{\nu b}-\omega_{\nu}{}^{ab}f_{\mu b}+2b_{[\mu}f_{\nu]}{}^{a}\nonumber \\
    & =T_{\mu\nu}^{(0)a}(f)+2b_{[\mu}f_{\nu]}{}^{a},\\
\label{curves}
R_{\mu\nu}&=\partial_{\mu}b_{\nu}-\partial_{\nu}b_{\mu}+4e_{[\mu}{}^{a}f_{\nu]a}, 
\end{align}
where $T_{\mu\nu}^{(0)}{}^{a}(e)$ and $R_{\mu\nu}^{(0)}{}^{ab}$ are the torsion and curvature component tensors of the 4-d Poincar\'e gravity.

As mentioned earlier, in our treatment, the resulting gauge group is determined by the SSB of the gauge group $SO(2,4)$. However, in addition, we have to guarantee the equivalence of the gauge transformations and diffeomorphisms. This requirement, as we shall see below, leads to the vanishing of the torsion and curvature tensors generalised in the present case as compared to those of the corresponding Poincar\'e case.

\subsection{Spontaneous symmetry breaking}
\label{sec4.1}
We shall start by choosing the parity conserving action, which is quadratic in terms of the field strength tensor \eqref{fst}, in which we have introduced a scalar that belongs in the adjoint rep, $\textbf{15}$, of $SU(4) \sim SO(6) \sim SO(2,4)$ along with a dimensionful parameter, $m$:
\begin{equation}
    S_{SO(2,4)}=a_{CG}\int d^4x [\operatorname{tr} \epsilon^{\mu \nu \rho \sigma} m\phi F_{\mu \nu}F_{\rho \sigma}+(\phi^2-m^{-2} \mathbb{1}_4)], 
\end{equation}
where the trace is defined as $ \operatorname{tr}\rightarrow \epsilon_{abcd} [\text{Generators}]^{abcd}$.

The scalar expanded on the generators is:
\begin{equation}
\phi=\phi^{a b} M_{a b}+\tilde{\phi}^a P_a+\phi^a K_a+\tilde{\phi} D\, .
\end{equation}

In accordance with \cite{Li:1973mq}, we pick the specific gauge in which $\phi$ is diagonal of the form $\operatorname{diag}(1,1,-1,-1)$. Specifically, we choose $\phi$ to be only in the direction of the dilatation generator $D$:
\begin{equation}
    \phi=\phi^0=\tilde{\phi}D \xrightarrow{\phi^2=m^{-2}\mathbb{1}_4}\phi=-2m^{-1} D.
\end{equation}
In this particular gauge, the action reduces to
\begin{equation}
    S=-2a_{CG}\int d^4x \operatorname{tr} \epsilon^{\mu \nu \rho \sigma} F_{\mu \nu}F_{\rho \sigma}D, 
\end{equation}
and the gauge fields $e,b$ and $\tilde{a}$ become scaled as $me,mb$ and $m\tilde{a}$ correspondingly. After straightforward calculations, using the expansion of the field strength tensor as in eq. \eqref{fst}, and the anticommutation relations of the generators, we obtain: 
\begin{equation}
\begin{gathered}
    S=-2a_{CG}\int d^4x \operatorname{tr} \epsilon^{\mu \nu \rho \sigma}\Big[\frac{1}{4}R_{\mu \nu}{}^{ab}R_{\rho \sigma}{}^{cd}M_{ab}M_{cd}D+\\
    +i\epsilon_{abcd}(R_{\mu \nu}{}^{ab}R_{\rho \sigma}{}^{c} K^d D - R_{\mu \nu}{}^{ab}\tilde{R}_{\rho \sigma}{}^{c}P^{d}D)+\\+(\frac{1}{2}\tilde{R}_{\mu \nu}{}^{a}R_{\rho \sigma} + 2\tilde{R}_{\mu \nu}{}^{a}R_{\rho \sigma}{}^{b})M_{ab}+\\
    +(\frac{1}{4}R_{\mu \nu}R_{\rho \sigma}- 2\tilde{R}_{\mu \nu}{}^{a}R_{\rho \sigma a})D   
    \Big].
\end{gathered}
\end{equation}
In this point we employ the trace on the several generators and their products. In particular:
\begin{equation}
\begin{gathered}
   \operatorname{tr}[K^{d}D]=\operatorname{tr}[P^{d}D]=\operatorname{tr}[M_{ab}]= \operatorname{tr}[D]=0, \\
        \text{and}\quad \operatorname{tr}[M_{ab}M_{cd}D]=-\frac{1}{2}\epsilon_{abcd}.
\end{gathered}
\end{equation}
The resulting broken action is:
\begin{equation}
\label{BrokenActionConformal}
     S_{\mathrm{SO}(1,3)}=\frac{a_{CG}}{4}\int d^4x \epsilon^{\mu \nu \rho \sigma}\epsilon_{abcd}R_{\mu \nu}{}^{ab}R_{\rho \sigma}{}^{cd},
\end{equation}
while its invariance has obviously been reduced only to Lorentz. Before continuing, we notice that there is no term containing the field $\tilde{a}_\mu$ present in the action. Thus, we may set $b_\mu=0$. This simplifies the form of the two component field strength tensors related to the $P$ and $K$ generators:
\begin{equation}
\begin{aligned}
  \tilde{R}_{\mu \nu}{}^a & =mT_{\mu \nu}^{(0) a}(e)-2 m^2b_{[\mu} e_{\nu]}{}^a \longrightarrow mT_{\mu \nu}^{(0) a}(e), \\
R_{\mu \nu}{}^a &=mT_{\mu \nu}^{(0) a}(f)+2m^2 b_{[\mu} f_{\nu]}{}^a \longrightarrow mT_{\mu \nu}^{(0) a}(f).
\end{aligned}
\end{equation}
The absence of the above field strength tensors in the action, allows us to also set $\tilde{R}_{\mu \nu}{}^a=R_{\mu \nu}{}^a=0$, and thus to obtain a torsion-free theory. Since $R_{\mu \nu}$ is also absent from the expression of the broken action, it may also be set equal to zero. From its definition in eq. \eqref{curves}, then we obtain the following relation among $e$ and $f$: 
\begin{equation}\label{ef}
    e_\mu{}^a f_{\nu a}-e_{\nu}{}^{a}f_{\mu a}=0.
\end{equation}
The above result reinforces one to consider solutions that relate $e$ and $f$. We examine two possible solutions of eq. \eqref{ef}.

\subsubsection*{When $f_\mu{}^a = ae_\mu{}^a$ - Einstein-Hilbert (E-H) action in the presence of a cosmological constant} In this case, first proposed in \cite{Chamseddine:2002fd}, by a simple substitution we obtain:
\begin{equation}
\begin{aligned}
        S_{\mathrm{SO}(1,3)} =\frac{a_{CG}}{4}\int d^4 x \epsilon^{\mu \nu \rho \sigma} \epsilon_{a b c d}&\left[R_{\mu \nu}^{(0) a b}-4m^2a\left(e_\mu{}^a e_\nu{}^b-e_\mu{}^b e_\nu{}^a\right)\right]\\
       &\left[R_{\rho \sigma}^{(0) c d}-4m^2a\left(e_\rho{}^c e_\sigma{}^d-e_\rho{}^d e_\sigma{}^c\right)\right]\longrightarrow\\\\
        S_{\mathrm{SO}(1,3)} =\frac{a_{CG}}{4}\int d^4 x \epsilon^{\mu \nu \rho \sigma} \epsilon_{a b c d}&[R_{\mu \nu}^{(0) a b}-8m^2ae_\mu{}^a e_\nu{}^b]\\
        &\qquad\qquad\left[R_{\rho \sigma}^{(0) c d}-8m^2ae_\rho{}^c e_\sigma{}^d\right],
       \end{aligned}
\end{equation}
which yields
\begin{equation}\label{so24finalaction}
\begin{aligned}
    S_{\mathrm{SO}(1,3)}=\frac{a_{CG}}{4}\int d^4 x \epsilon^{\mu \nu \rho \sigma} \epsilon_{a b c d}[R_{\mu \nu}^{(0) a b}R_{\rho \sigma}^{(0) c d}-16m^2aR_{\mu \nu}^{(0) a b}e_\rho{}^c e_\sigma{}^d+\\
    +64m^4a^2 e_\mu{}^a e_\nu{}^b e_\rho{}^c e_\sigma{}^d].
\end{aligned}
\end{equation}

This action consists of three terms: a Gauss-Bonnet topological term, the E-H action, and a cosmological constant term. For $a<0$, the above describes GR in AdS space.

\subsubsection*{When $f_\mu{}^a = -\frac{1}{4} (R_\mu{}^a - \frac{1}{6} R e_\mu{}^a)$ - Weyl action}
This relation between $f$ and $e$ was suggested in refs \cite{Kaku:1978nz} and \cite{freedman_vanproeyen_2012}. It can be obtained as a solution of \eqref{ef}, by solving the constraint wrt $f_{\mu}{}^a$. In particular:
\begin{align}\label{re}
    R_{\mu\nu}{}^{ab}e_b{}^\nu=0 \xRightarrow{\eqref{Rmnab}} R_{\mu\nu}^{(0)ab}e^\nu{}_b-8e_{[\mu}{}^{[a}f_{\nu]}{}^{b]}=0.
\end{align}
The Weyl tensor can be expressed in terms of the Riemann tensor as \cite{freedman_vanproeyen_2012}:
\begin{align}
C_{\mu \nu a b}=R_{\mu \nu a b}+\frac{1}{n-2}\left(g_{\nu a} R_{\mu b}+g_{\mu b} R_{\nu 
 a}-g_{\mu a} R_{\nu b}-g_{\nu b} R_{\mu a}\right)+\\
 +\frac{1}{(n-1)(n-2)} R( g_{\mu a} g_{\nu b}-g_{\mu b} g_{\nu a}).
\end{align}
Writing the above equality in terms of the Riemann tensor and replacing it into \eqref{re}, we obtain:
\begin{equation}
\frac{1}{(n-1)(n-2)} e_\mu{}^a e_b{}^\nu R_\nu{}^b-R_{\mu \nu}^{(0) a b} e^\nu{}_b=4 f_\mu{}^a,
\end{equation}
where the Weyl tensor contractions have vanished, since it is antisymmetric.\\
For $n=4$ the above expression equals:
\begin{equation}
\frac{1}{6} e_\mu{}^a e_b{}^\nu R_\nu{}^b-R_{\mu \nu}^{(0) a b} e^\nu{}_b=4 f_\mu{}^a,
\end{equation}
Employing the contractions with the Riemann tensor,  $R_\mu{}^a=R_{\mu \nu}{}^{(0) a b} e_b{}^\nu$ and $R=e_a{}^\mu R_\mu{}^a$, and solving in terms of $f_{\mu}{}^a$, one obtains
\begin{equation}
\label{f.e.relation}
f_\mu{}^a=-\frac{1}{4}\left(R_\mu{}^a-\frac{1}{6} R e_\mu{}^a\right).
\end{equation}
Taking this into account we obtain the following action:
\begin{equation}
    \begin{aligned}
             S_W=\frac{a_{CG}}{4}\int d^4 x \epsilon^{\mu \nu \rho \sigma} \epsilon_{a b c d}\\
             &\left[R_{\mu \nu}^{(0) a b}+\frac{1}{2}\left(m e_\mu{}^{[a} R_\nu{}^{b]}-me_\nu{}^{[a} R_\mu{}^{b]}\right)-\frac{1}{3} m^2 R e_\mu{}^{[a} e_\nu{}^{b]}\right]\\
             &\left[R_{\rho \sigma}^{(0) c d}+\frac{1}{2}\left(m e_\rho{}^{[c} R_\sigma{}^{d]}-m e_\sigma{}^{[c} R_\rho{}^{d]}\right)-\frac{1}{3} m^2 R e_\rho{}^{[c} e_\sigma{}^{d]}\right].
    \end{aligned}
\end{equation}
Considering the rescaled vierbein $\tilde{e}_\mu{}^{a}=m e_\mu{}^{a}$ and recalling that $R_{\mu \nu}^{(0) a b}=-R_{\nu \mu}^{(0) a b}$, we obtain
\begin{equation}
    \begin{aligned}
             S_W=\frac{a_{CG}}{4}\int d^4 x&\epsilon^{\mu \nu \rho \sigma} \epsilon_{a b c d}\\&\left[R_{\mu \nu}^{(0) a b}-\frac{1}{2}\left(\tilde{e}_\mu{}^{[a} R_\nu{}^{b]}-\tilde{e}_\nu{}^{[a} R_\mu{}^{b]}\right)+\frac{1}{3} R \tilde{e}_\mu{}^{[a} \tilde{e}_\nu{}^{b]}\right]\\
             &\left[R_{\rho \sigma}^{(0) c d}-\frac{1}{2}\left( \tilde{e}_\rho{}^{[c} R_\sigma{}^{d]}- \tilde{e}_\sigma{}^{[c} R_\rho{}^{d]}\right)+\frac{1}{3} R \tilde{e}_\rho{}^{[c} \tilde{e}_\sigma{}^{d]}\right],
    \end{aligned}
\end{equation}
which is equal to 
\begin{align}
\label{weyl1}
             S_W=\frac{a_{CG}}{4}\int d^4 x \epsilon^{\mu \nu \rho \sigma} \epsilon_{a b c d}C_{\mu \nu}{}^{a b}C_{\rho \sigma}{}^{c d},
\end{align}
where $C_{\mu \nu}{}^{a b}$ is the Weyl conformal tensor. This action leads to the well-known four-dimensional scale invariant Weyl action,
\begin{equation}
\label{weyl2}
S_W =2a_{CG}\int \mathrm{d}^4 x\left(R_{\mu \nu} R^{\nu \mu}-\frac{1}{3} R^2\right).
\end{equation}

The Weyl action of WG in the forms given in eqs. \eqref{weyl1} and \eqref{weyl2}, being scale invariant, naturally does not contain a cosmological constant. WG is an attractive possibility for describing gravity at high scales (for some recent developments see \cite{Maldacena:2011mk, mannheim, Anastasiou:2016jix, ghilencea2023, Hell:2023rbf, Condeescu:2023izl}, as CG does. However, in the case that WG is obtained after the SSB of the CG, as is described above, a question remains on how one can obtain Einstein gravity from the SSB of WG. Our suggestion is the following. We start again from CG and we introduce a scalar in the \textbf{15} rep of $SU(4) \sim SO(6) \sim SO(2,4)$ as described above and by choosing relation \eqref{f.e.relation}, we are led after the SSB of the scalar $\textbf{15}$-plet to the Weyl action. In addition, we introduce a scalar in the 2nd rank anti-symmetric tensor of $SU(4)$, \textbf{6}, which after SSB leads to the E-H action. One can easily see the result of these breakings by considering the decompositions of the \textbf{15} generators of $SU(4)$ under $SU(2) \times SU(2) \times U(1)$, describing the gauge group of the Lorentz group and dilatations to which $SU(4)$ breaks after the SSB due to the scalar $\textbf{15}$-plet,
\begin{align}
\label{SU4ToSU2SU2U1}
    SU(4)&\xrightarrow{<\textbf{15}>}SU(2) \times SU(2) \times U(1) \nonumber\\
    \textbf{15}&=[(\textbf{3},\textbf{1})_0+(\textbf{1},\textbf{3})_0]+(\textbf{1},\textbf{1})_0+(\textbf{2},\textbf{2})_2+(\textbf{2},\textbf{2})_2\,,
\end{align}
where the $[(\textbf{3},\textbf{1})_0+(\textbf{1},\textbf{3})_0]$ describes the generators of the Lorentz gauge group, $M_{ab}$, the $(\textbf{1},\textbf{1})_0$ the generator of dilatations, $D$, the $(\textbf{2},\textbf{2})_2$ the generators of the translations, $P_a$ and the $(\textbf{2},\textbf{2})_2$ the generators of conformal transformations, $K_a$ (there is an arbitrariness in the choice of the last two sets of generators). The generators $P_a$ and $K_a$ are broken due to the SSB of the scalar $\textbf{15}$-plet. Similarly the decompositon of the 15 generators of $SU(4)$ under the $SO(5)$ to which it breaks after the SSB of the scalar $\textbf{6}$-plet is,
\begin{align}
\label{SU4ToSO5}
    SU(4)&\xrightarrow{<\textbf{6}>}SO(5) \nonumber\\
    \textbf{15}&=\textbf{10}+\textbf{5}\,,
\end{align}
where the \textbf{10} describes the generators of the unbroken gauge group $SO(5)$ and \textbf{5} the broken generators. To identify the unbroken and the broken generators in \eqref{SU4ToSO5} it helps to consider the decomposition of reps \textbf{10} and \textbf{5} of $SO(5)$ under the $SU(2) \times SU(2)$ describing the Lorentz gauge group in \eqref{SU4ToSU2SU2U1},
\begin{align}
    SO(5)& \supset SU(2) \times SU(2) \nonumber\\
    \textbf{10}&=(\textbf{3},\textbf{1})+(\textbf{1},\textbf{3})+(\textbf{1},\textbf{1})+(\textbf{2},\textbf{2})\,,\\
    \textbf{5}&=(\textbf{1},\textbf{1})+(\textbf{2},\textbf{2})\,.
\end{align}
Now one can easily recognize the ten unbroken generators from the SSB of the scalar \textbf{6}-plet correspond to the Lorentz generators, $M_{ab}$ and the generators of the translations, $P_a$ (which, though they were broken by the $<$\textbf{15}$>$), while the five broken generators can be identified with the generators $(\textbf{1},\textbf{1})$ of dilatations and the $(\textbf{2},\textbf{2})$ of $K_a$.

In summary, $<$\textbf{15}$>$ breaks the generators of $P_a$ and $K_a$, leaving unbroken the Lorentz rotation generators, $M_{ab}$ and the dilaton generator, $D$, while $<$\textbf{6}$>$ breaks the dilaton generator, $D$ and gives an additional contribution to the breaking of the generators $K_a$ (and to the masses of the corresponding gauge bosons).

\subsection{Gauge and diffeomorphisms transformations}
Following the same procedure as in the 3-d case, in Section \ref{sec2}, we calculate the difference between a diffeomorphism and a gauge transformation of the fields. Specifically, $\tilde{\delta}e_\mu{}^a-\delta e_\mu{}^a$ is considered using \eqref{eGCT} and \eqref{deltaee}:
\begin{equation}
\tilde{\delta}e_\mu^a-\delta e_\mu^a=\left(v^\nu\partial_\nu e_\mu^a+\partial_\mu(v^\nu e_\nu^a)-v^\nu\partial_\mu e_\nu^a\right)-\left(\partial_\mu\xi^a+\omega_{\mu}{}^a{}_b\xi^b-b_\mu\xi^a-\lambda^{ab}e_\mu^b+\kappa e_\mu^a\right).
\end{equation}
Then setting $\xi^a=v^\mu e_\mu{}^a$, $\lambda^{ab}=v^\mu \omega_\mu{}^{ab}$, $\kappa=v^\mu b_\mu$, and $\rho^a=v^\mu f_\mu{}^a$, the above difference takes the following form,
\begin{equation}
\tilde{\delta}e_{\mu}{}^{a}-\delta e_{\mu}{}^{a}=v^{\nu}\left(\partial_{\nu}e_{\mu}{}^{a}-\partial_{\mu}e_{\nu}{}^{a}-\omega_{\mu}{}^{a}{}_{b}e_{\nu}{}^{b}+\omega_{\nu}{}^{a}{}_{b}e_{\mu}{}^b+b_{\mu}e_{\nu}{}^{a}-b_{\nu}e_{\mu}{}^{a}\right)=-v^{\nu}\tilde{R}_{\mu\nu}{}^{a}.
\end{equation}
Clearly the constraint that is needed for getting rid of the translational part of the theory, with a coordinate transformation making up for them, is the vanishing of the torsion,
\begin{equation}
    \tilde{R}_{\mu\nu}{}^a=0\,.
\end{equation}

Similarly, the difference between a diffeomorphism and the gauge transformation $\tilde{\delta}f_\mu{}^a-\delta f_\mu{}^a$ leads to
\begin{equation}
    R_{\mu\nu}{}^a=0\, ,
\end{equation}
while the corresponding difference $\tilde{\delta}\omega_\mu{}^{ab}-\delta \omega_\mu{}^{ab}$ results to
\begin{equation}
    R_{\mu\nu}{}^{ab}=0\, .
\end{equation}
As already mentioned in Section \ref{sec4.1}, the generators $P_a$ and $K_a$ are broken due to the SSB of the scalar 15-plet, i.e. the two torsionless conditions are resulting from the SSB of the scalar 15-plet.

Therefore the two torsionless conditions and the vanishing of the curvature tensor, which is satisfied on-shell guarantee the equivalence of the diffeomorphisms and gauge transformations. In other words, the gauge theory based on the $SO(2,4)$ group describes the 4-d conformal gravity.

At this point, it should be noted that eq. \eqref{re}, which eventually leads to WG, is an additional constraint and is not necessary in order to guarantee the equivalence of the diffeomorphisms and gauge transformations.

\section{Noncommutative Gauge Theory of 4D Gravity - Fuzzy Gravity}
\label{sec5}

\subsection{Gauge theories on noncommutative spaces}
In this section we include the basics regarding the construction of gauge theories on noncommutative spaces, since it is fundamental for our purposes. In noncommutative geometry, gauge fields arise in a very natural way and are intertwined with the notion of covariant coordinate \cite{Madore:2000en}, that is the noncommutative analogue of the covariant derivative as we will stress later. Let us now begin with considering a field $\phi(X_a)$ on a fuzzy space, depending on the noncommuting coordinates $X_{a}.$ The field belongs to a representation of a gauge group $G$, therefore an infinitesimal gauge transformation $\delta\phi$ with gauge transformation parameter $\lambda(X_a)$ is given by:
\begin{equation}
    \delta\phi(X)=\lambda(X)\phi(X)\,.
\end{equation}
In case the transformation parameter $\lambda(X)$ is simply a function of the coordinates, $X_\alpha$, then it is considered as an infinitesimal Abelian transformation and the gauge group is $G=U(1)$, while in case $\lambda(X)$ is a $P\times P$ matrix, then it can be viewed as a gauge transformation of the non-Abelian gauge group $G=U(P)$, i.e. the group including all hermitian $P\times P$ matrices. It is worth noting that the coordinates are invariant under transformations of the gauge group, $G$, that is $\delta X_\alpha=0$. In turn, let us perform a gauge transformation on the product of a coordinate and the field:
\begin{equation}
    \delta(X_a\phi)=X_a\lambda(X)\phi\, ,
\end{equation}
The above transformation is not a covariant one since, in general:
\begin{equation}
    X_{a}\lambda(X)\phi\neq\lambda(X)X_{a}\phi\,.
\end{equation}
Drawing ideas from the methodology of ordinary gauge theories, in which covariant derivative is defined for similar reasons, in the noncommutative case, the covariant coordinate, $\phi_{\alpha}$, is introduced by its transformation rule:
\begin{equation}
\label{3.5}
    \delta(\phi_a\phi)=\lambda\phi_a\phi\,,
\end{equation}
which is satisfied in case
\begin{equation}
\label{3.6}
    \delta(\phi_a)=[\lambda,\phi_a]\,.
\end{equation}
Eventually, the covariant coordinate is defined as:
\begin{equation}
    \phi_a\equiv X_a+A_a\,,
\end{equation}
where it is straightforward to identify $A_\alpha$ as the gauge connection of the theory. Putting together equations \eqref{3.5} and \eqref{3.6}, the gauge transformation of the connection, $A_{\alpha}$, is obtained:
\begin{equation}
    \delta A_a=-[X_a,\lambda]+[\lambda,A_a]\,,
\end{equation}
giving an a posteriori explanation of the interpretation of $A_a$ as a gauge field (for more details, see \cite{Aschieri:2004vh}). Accordingly, the corresponding field strength tensor, $F_{ab}$ , is defined as follows:
\begin{equation}
    F_{ab} \equiv [X_a, A_b] - [X_b, A_a] + [A_a, A_b] - C_{ab}{}^c A_c = [\phi_a, \phi_b] - C_{ab}{}^c \phi_c,
\end{equation}
which is easily proven to be covariant under a gauge transformation,
\begin{equation}
    \delta F_{ab} = [\lambda, F_{ab}].
\end{equation}
The above scheme will be used in the following sections in the construction of gravity model as gauge theory on fuzzy covariant space.

\subsection{The Background Space}
The next step we need to take before we move on with the gauge theory of FG, is to establish the background space, on which this theory will be formulated. In refs \cite{yang1947, Heckman_2015,Manolakos_paper1, Manolakos_paper2, Manolakos:2022universe} extending the original Snyder's suggestion \cite{Snyder:1946qz} the authors have considered the group the $SO(1,5)$ and have assigned the 4-d spacetime coordinates to elements of its Lie algebra.

More specifically starting with the group $SO(1,5)$, whose generators obey the following Lie algebra:
\begin{equation}
     \left[J_{mn},J_{rs}\right]=i\left(\eta_{mr}J_{ns}+\eta_{ns}J_{mr}-\eta_{nr}J_{ms}-\eta_{ms}J_{nr}\right),
\end{equation}
where $m,n,r,r = 0,\ldots, 5$, and $\eta_{mn} = diag(-1, 1, 1, 1, 1, 1)$. Performing the decompositions of $SO(1, 5)$ to its maximal subgroups, up to $SO(1, 3)$, i.e., $SO(1,5) \supset  SO(1, 4)$ and $SO(1, 4) \supset SO(1, 3)$, turns the above commutation relation to the following:
\begin{equation}
\begin{gathered}  
    \left[J_{ij},J_{kl}\right]=i\left(\eta_{i k}J_{j l}+\eta_{j l}J_{i k}-\eta_{j k}J_{i l}-\eta_{i l}J_{j k}\right),\\
    \left[J_{i j},J_{k5}\right]=i\left(\eta_{i k}J_{j5}-\eta_{j k}J_{i5}\right),\\
    \left[J_{i5},J_{j5}\right]=i J_{ij},\\
    \left[J_{i j},J_{k4}\right]=i\left(\eta_{i k}J_{j4}-\eta_{j k}J_{i4}\right),\\
    \left[J_{i4},J_{j4}\right]=i J_{ij},\\
    \left[J_{i4},J_{j5}\right]=i \eta_{ij}J_{45},\\
    \left[J_{i j},J_{45}\right]=0,\\
    \left[J_{i 4},J_{45}\right]=-i J_{i5},\\
    \left[J_{i 5},J_{45}\right]=i J_{i4}.
\end{gathered}
\end{equation}
Next one may convert the generators to physical quantities by setting
  \begin{equation}
    \Theta_{ij}=\hbar J_{ij}, \, \text{and} \ X_i=\lambda J_{i5},
\end{equation}
where $\lambda$ is a natural unit of length, and furthermore identify the momenta as
\begin{equation}
\label{identifications.P}
    P_i=\frac{\hbar}{\lambda}J_{i4},
\end{equation}
and set $h=J_{45}$. Then given these identifications and the commutation relations above, one obtains:
\begin{equation}
\begin{gathered}  
    \left[\Theta_{ij},\Theta_{kl}\right]=i \hbar\left(\eta_{i k}\Theta_{j l}+\eta_{j l}\Theta_{i k}-\eta_{j k}\Theta_{i l}-\eta_{i l}\Theta_{j k}\right),\\
    \left[\Theta_{i j},X_{k}\right]=i\hbar \left(\eta_{i k}X_{j}-\eta_{j k}X_{i}\right),\\
    \left[\Theta_{i j},P_{k}\right]=i\hbar\left(\eta_{i k}P_{j}-\eta_{j k}P_{i}\right),\\
    \left[X_{i},X_{j}\right]=\frac{i\lambda^2}{\hbar} \Theta_{ij},\ 
    \left[X_{i},P_{j}\right]=i\hbar \eta_{ij}h,\ 
    \left[P_{i},P_{j}\right]=\frac{i \hbar}{\lambda^2} \Theta_{ij},\\
    \left[X_{i},h\right]=\frac{i\lambda^2}{\hbar} P_{i},\ 
    \left[P_{i},h\right]=-\frac{i\hbar}{\lambda^2} X_{i},\ 
    \left[\Theta_{i j},h\right]=0.
\end{gathered}
\end{equation}
From the above relations, it becomes clear that one is led to the following significant results. First since the coordinates as well as the momenta are elements of this Lie algebra, they exhibit a noncommutative behavior, implying that both the space-time and the momentum space become quantized. In addition it becomes evident that the commutation relation between coordinates and momenta naturally yields a Heisenberg-type uncertainty relation.

\subsection{Gauge Group and Representation}
Starting with the formulation of a gauge theory for gravity in the space mentioned above, the first step is to determine the group that shall be gauged. Naturally, the group we consider is the one that describes the symmetries of the theory, in this case, the isometry group of $dS_4$, $SO(1,4)$. It is known though, that in the particular case of gauge theories built on noncommutative spaces, on top of the commutators of the various fields, we also have to properly treat their anticommutators. Specifically, let us consider two elements of an arbitrary algebra, $\varepsilon(X)=\varepsilon^{a}(X) T_{a}$ and $\phi(X)=\phi^{a}(X) T_{a}$, where $T_a$ are the generators of the algebra. Their commutation relation will be:
\begin{equation}
\label{anticom}
[\varepsilon, \phi]=\frac{1}{2}\left\{\varepsilon^{a}, \phi^{b}\right\}\left[T_{a}, T_{b}\right]+\frac{1}{2}\left[\varepsilon^{a}, \phi^{b}\right]\left\{T_{a}, T_{b}\right\}.
\end{equation}
In the ordinary, commutative case, the last term vanishes as the components $\varepsilon^a$ and $\phi^b$ are ordinary functions of coordinates which commute with each other. On the contrary, when the space is noncommutative, the last term is not vanishing and, thus, the anticommutator of the generators remains in the expression. In the general case, an anticommutator like that will yield items that don't belong in the original algebra of the theory, meaning that the corresponding algebra does not close. In order to remedy that, we have to pick a specific representation of the algebra generators, and subsequently extend the initial gauge group to one with larger symmetry, in which both the commutator and anticommutator algebras close. Following this procedure, we are led to the extension of our initial gauge group $SO(1,4)$ to $SO(2,4) \times U(1)$.

\subsection{Fuzzy Gravity}

Since we have already determined the appropriate gauge group of the theory, we are now able to move on with the gauging procedure on the fuzzy space that was presented above. Following the steps described in \cite{Manolakos_paper1}, we firstly have to introduce the covariant coordinate of the theory, which is defined as:
\begin{equation}\label{CovariantCoordinate}
    \mathcal{X}_\mu=X_\mu \otimes \mathbb{1}_4 +A_\mu (X)\, ,
\end{equation}
where $A_\mu$ is the gauge connection of the theory. The gauge connection can, in turn, be expanded on the gauge group generators as:
\begin{equation}\label{GaugeConnectionFuzzy}
    A_\mu = a_\mu \otimes \mathbb{1}_{4} +  \omega_\mu{}^{ab}\otimes M_{ab}+ e_\mu{}^{a}\otimes P_a + f_\mu{}^{a}\otimes K_a + \tilde{a}_\mu\otimes D\,.
\end{equation}
Given the above expansion \eqref{GaugeConnectionFuzzy}, the explicit form of the covariant coordinate \eqref{CovariantCoordinate} can be written down as:
\begin{equation}
    \mathcal{X}_\mu= (X_\mu + a_\mu) \otimes \mathbb{1}_{4} +  \omega_\mu{}^{ab}\otimes M_{ab}+ e_\mu{}^{a}\otimes P_a + f_\mu{}^{a}\otimes K_a + \tilde{a}_\mu\otimes D\,.
\end{equation}
At this point, what remains to be determined is the appropriate covariant field strength tensor for the theory. In the case of noncommutativity, the latter is defined as \cite{Madore_1992, Manolakos_paper1}:
\begin{equation}
    \hat{F}_{\mu \nu}\equiv\left[\mathcal{X}_{\mu}, \mathcal{X}_{\nu}\right]-\kappa^2 \hat{\Theta}_{\mu \nu}\, ,
\end{equation}
where $\hat{\Theta}_{{\mu}{\nu}}\equiv\Theta_{{\mu}{\nu}}+\mathcal{B}_{{\mu}{\nu}}$, in which $\mathcal{B}_{{\mu}{\nu}}$ is a 2-form field taking care of the transformation of $\Theta$, promoting it to its covariant form. Since $\hat{F}_{\mu \nu}$ is an element of the gauge algebra it can also be expanded on the algebra's generators as 
\begin{equation}
    \hat{F}_{\mu \nu}= R_{\mu \nu} \otimes \mathbb{1}_4 +\frac{1}{2} R_{\mu \nu}{}^{a b} \otimes M_{a b} + \tilde{R}_{\mu \nu}{}^{a} \otimes P_{a}+R_{\mu \nu}{}^{a} \otimes K_{a}+\tilde{R}_{\mu \nu} \otimes D\,.
\end{equation}
The SSB goes along the same lines as the one described in the case of Conformal Gravity, i.e. we introduce a scalar field, $\Phi(X)$, belonging to the 2nd rank antisymmetric rep of $SO(2,4)$, in the action and fix it in the gauge that leads to the Lorentz group (see \cite{Manolakos_paper1, Manolakos_paper2, Roumelioti:2024lvn}). This scalar field must be charged under the $U(1)$ gauge symmetry so that the $U(1)$ part breaks and it doesn't appear in the final symmetry. Introducing the scalar field, the action takes the form:
\begin{equation}
\mathcal{S}=\operatorname{Trtr} \Big[\lambda \Phi(X) \varepsilon^{\mu \nu \rho \sigma}\hat{F}_{\mu \nu} \hat{F}_{\rho \sigma} +\eta\left(\Phi(X)^2-\lambda^{-2} \mathbb{1}_N \otimes \mathbb{1}_4\right)\Big],
\end{equation}
where the first trace is over the coordinate matrices, the second is over the generators of the gauge group, $\eta$ is a Lagrange multiplier, and $\lambda$ is a dimensionfull parameter. The scalar field itself is also an element of the gauge group, and hence can be expanded on its generators, as
\begin{equation}
\begin{aligned}
\Phi(X)=\phi(X) &\otimes \mathbb{1}_4+\phi^{a b}(X) \otimes M_{a b}+\tilde{\phi}^a(X) \otimes P_a+\\&+\phi^a(X) \otimes K_a+\tilde{\phi}(X) \otimes D.
\end{aligned}
\end{equation}
As mentioned above, just like in our previous works  \cite{Manolakos_paper1, Manolakos_paper2, Roumelioti:2024lvn}, we gauge-fix the scalar field into the dilatation direction:
\begin{equation}
\Phi(X)=\left.\tilde{\phi}(X) \otimes D\right|_{\tilde{\phi}=-2 \lambda^{-1}}=-2 \lambda^{-1} \mathbb{1}_N \otimes D.
\end{equation}
On-shell, when the above equation holds, the aforementioned action reduces to the following form:
\begin{equation}
\mathcal{S}_{b r}=\operatorname{Tr}\left(\frac{\sqrt{2}}{4} \varepsilon_{a b c d} R_{m n}{}^{a b} R_{r s}{}^{c d}-4 R_{m n} \tilde{R}_{r s}\right) \varepsilon^{m n r s},
\end{equation}
while any other term, along with the Lagrange multiplier, has vanished due to the gauge fixing. This resulting action now bears the remaining $SO(1,3)$ gauge symmetry, following the SSB. Moreover, when the commutative limit of the above action is considered (for details, see \cite{Manolakos_paper2}), it reduces to the Palatini action, which in turn is equivalent to EG, with a cosmological constant term present.

\section{Unification of Conformal and Fuzzy Gravities with
Internal Interactions, Fermions and Breakings}
\label{sec6}

According to the suggestion in \cite{Roumelioti:2024lvn} the unification of the CG with internal interactions described by the $SO(10)$ could be achieved using the $SO(2,16)$ as the unifying gauge group. As it was emphasized in Section \ref{sec1} the whole strategy was based on the observation that the dimension of the tangent space is not necessarily equal to the dimension of the corresponding curved manifold \cite{roumelioti2407, Percacci:1984ai, Percacci_1991, Nesti_2008, Nesti_2010, Krasnov:2017epi, Chamseddine2010, Chamseddine2016, noncomtomos, Konitopoulos:2023wst, Weinberg:1984ke}. An additional fundamental observation \cite{Roumelioti:2024lvn} is that in the case of $SO(2,16)$ one can impose Weyl and Majorana conditions on fermions \cite{D_Auria_2001, majoranaspinors}. CG is obtained by gauging the $SO(2,4) \sim SU(2,2) \sim SO(6) \sim SU(4)$  (the last two in Euclidean signature). Therefore, starting with the gauge group $SO(2,16)$ we obtain first $C_{SO(2,16)} (SO(2,4)) = SO(12)$, which should break further to give us the $SO(10)$ that will be used to describe the internal interactions.

More specifically, using Euclidean signature for simplicity (the implications of using non-compact space are explicitly discussed in \cite{Roumelioti:2024lvn} and below in Section \ref{sec6.1}), one starts with $SO(18)$ and with the fermions in its spinor representation, \textbf{256}. Then the SSB of $SO(18)$ leads to its maximal subgroup $SO(6) \times SO(12)$. Let us recall for convenience the branching rules of the relevant reps \cite{Slansky:1981yr, Feger_2020, Li:1973mq},
\begin{equation}\label{so18}
\begin{aligned}
SO(18) & \supset S O(6) \times S O(12) & & \\
{\textbf{18}} & =(\textbf{6},\textbf{1}) + (\textbf{1}, \textbf{12})  & & \text { vector } \\
{\textbf{153}} & =(\textbf{15}, \textbf{1}) + (\textbf{6}, \textbf{12}) + (\textbf{1}, \textbf{66}) & & \text { adjoint } \\
{\textbf{256}} & =({\textbf{4}}, \overline{{\textbf{32}}})+(\overline{{\textbf{4}}}, {\textbf{32}}) & & \text { spinor } \\
{\textbf{170}} & =({\textbf{1}}, {\textbf{1}})+({\textbf{6}}, {\textbf{12}})+\left({\textbf{2 0}}^{\prime}, {\textbf{1}}\right)+({\textbf{1}}, {\textbf{77}}) & & \text { 2nd rank symmetric }
\end{aligned}  
\end{equation}

The SSB of $SO(18)$ to $SO(6) \times SO(12)$ is done by giving a vev to the $<$\textbf{1},\textbf{1}$>$ component of a scalar in the \textbf{170} rep. Concerning fermions, we start with the spinor rep, \textbf{256}. However, since the Majorana condition can be imposed, due to the non-compactness of the used $SO(2,16) \sim SO(18)$, after the SSB we are led to the $SO(6) \times SO(12)$ gauge theory with fermions in the $(\overline{\textbf{4}}, \textbf{32})$  representation\footnote{for details see Section \ref{sec6.1}}.

In order to further break $SO(12)$ down to $SO(10) \times U(1)$ or to $SO(10) \times U(1)_{\text{global}}$ we can use scalars either in rep \textbf{66} (contained in the adjoint, \textbf{153}, of $SO(18)$) or in the \textbf{77} (contained in the \textbf{170} of $SO(18)$) respectively, given the branching rules,

\begin{equation}
\begin{aligned}
SO(12) &\supset S O(10) \times U(1)\\
\textbf{66} &= (\textbf{1})(0) + (\textbf{10})(2) + (\textbf{10})(-2) + (\textbf{45})(0)\\
\textbf{77} &= (\textbf{1})(4) + (\textbf{1})(0) + (\textbf{1})(-4)+ (\textbf{10})(2) + (\textbf{10})(-2) + (\textbf{54})(0)\, .
\end{aligned}
\end{equation}
According to the above, by giving vev to the $<$(\textbf{1})(0)$>$ of the \textbf{66} rep we obtain the gauge group $SO(10) \times U(1)$ after the SSB, and by giving vev to the $<$(\textbf{1})(4)$>$ of the \textbf{77} rep we obtain $SO(10) \times U(1)_{\text{global}}$ after the SSB. 

Similarly, we can further break $SU(4)$ down to $SO(4) \sim SU(2) \times SU(2)$ in two steps. First, we break it to $SO(2,3) \sim SO(5)$, and then to $SO(4)$. For that, recall the following branching rules \cite{Krasnov:2017epi, Chamseddine2016}:
\begin{equation}
\begin{aligned}
SU(4) &\supset SO(5)\\
\textbf{4} &= \textbf{4}\\
\textbf{6} &= \textbf{1}+\textbf{5}\, .
\end{aligned}
\end{equation}

As a first step by giving vev to a scalar that belongs in rep \textbf{6} of $SU(4)$ (which belongs in the \textbf{18} rep of $SO(18)$) in the $<$\textbf{1}$>$ component, the $SU(4)$ breaks down to the $SO(5)$. Then according to the branching rules:
\begin{equation}
\begin{aligned}
SO(5) &\supset SU(2)\times SU(2)\\
\textbf{5} &= (\textbf{1},\textbf{1}) + (\textbf{2},\textbf{2})\\
\textbf{4} &= (\textbf{2},\textbf{1}) + (\textbf{1},\textbf{2})\, ,
\end{aligned}
\end{equation}
by giving vev in $<$\textbf{1}, \textbf{1}$>$ of a scalar in the \textbf{5} rep of $SO(5)$ (contained in the \textbf{6} of $SU(4)$) and in the \textbf{18} of $SO(18)$), we eventually obtain the Lorentz group $SU(2) \times SU(2) \sim SO(4) \sim SO(1,3)$. Note in addition, that in this case the rep \textbf{4} is decomposed under $SU(2) \times SU(2) \sim SO(1,3)$ in the appropriate reps to describe the two Weyl spinors.

One can follow also another way to break $SU(4)$ to $SU(2) \times SU(2)$, along the lines discussed in Section \ref{sec4.1}, i.e. in order to break the $SU(4)$ gauge group to $SU(2) \times SU(2)$ we can use scalars in the adjoint rep of $SU(4)$, \textbf{15}, which is contained in the adjoint rep of $SO(18)$, \textbf{153}. In that case we have:
\begin{equation}
\begin{aligned}
SU(4) &\supset SU (2) \times SU (2) \times U (1)\\
\textbf{4} &= (\textbf{2},\textbf{1})(1) + (\textbf{1},\textbf{2})(-1)\\
\textbf{15} &= (\textbf{1}, \textbf{1})(0) + (\textbf{2}, \textbf{2})(2) + (\textbf{2}, \textbf{2})(-2)+ (\textbf{3}, \textbf{1})(0) + (\textbf{1}, \textbf{3})(0)\, .
\end{aligned}
\end{equation}
Then, by giving vev in the $<$\textbf{1}, \textbf{1}$>$ direction of the adjoint rep, \textbf{15} we obtain the known result \cite{Li:1973mq}, that $SU(4)$ breaks spontaneously to $SU(2) \times SU(2) \times U(1)$. The way to vanish the corresponding $U(1)$ gauge boson and remain with the $SU(2) \times SU(2)$ was discussed already in Section \ref{sec4.1}. Note again that the \textbf{4} is decomposed in the  appropriate reps of $SU(2) \times SU(2) \sim SO(1,3)$ able to describe the two Weyl spinors.

Having established the analysis of the various breakings using the branching rules under the maximal subgroups starting from the group $SO(18)$, one can easily consider instead the isomorphic algebras of the various groups. Specifically, instead of $SO(18)$, the isomorphic algebra of the non-compact groups $SO(2,16) \sim SO(18)$, and similarly $SO(2,4) \sim SO(6) \sim SU(4)$.

\subsection{Weyl and Majorana conditions on Fermions}
\label{sec6.1}
Having examined the various SSB in Section \ref{sec4.1} and above, let us next discuss further the fermions and the result of Weyl and Majorana conditions when they are imposed on them.

A Dirac spinor, $\psi$ has $2^{\frac{D}{2}}$ independent components in D dimensions. Then the Weyl and Majorana conditions, when imposed, each divide the number of independent components by 2. The Weyl constraint can be imposed only for even D, therefore the Weyl–Majorana spinor, resulting after the imposition of both Weyl and Majorana conditions to a Dirac spinor, has $2^{\frac{D-4}{2}}$ independent components (for even D). 

The unitary reps of the Lorentz group $SO(1,D-1)$ are labeled by a continuous momentum vector k, and by a spin `projection', which in D dimensions is a rep of the compact subgroup $SO(D-2)$. The Dirac, Weyl, Majorana, and Weyl–Majorana spinors carry indices that transform as finite-dimensional non-unitary spinor reps of $SO(1, D-1)$. It is also known \cite{D_Auria_2001, majoranaspinors}, that the type of spinors one obtains for $SO(p,q)$ in the real case is governed by the signature $(p-q)\ \text{mod}\ 8$. Among even signatures, signature zero gives a real rep, signature four a quaternionic rep, while signatures two and six give complex reps. In the case of $SO(2,16)$ the signature is six, and imposing the Majorana condition in addition to Weyl is permitted.

For completeness and fixing the notation let us recall, the well-known case of 4 dimensions, which were discussed only briefly earlier in the present section. The $SO(1,3)$ spinors in the usual $SU(2) \times SU(2)$ basis transform as $(\textbf{2}, \textbf{1})$ and $(\textbf{1}, \textbf{2})$, with reps labelled by their dimensionality. The two-component Weyl spinors, $\psi_L$ and $\psi_R$, transform as the irreducible spinors, $\psi_L \sim (\textbf{2},\textbf{1})$ and $\psi_R \sim (\textbf{1},\textbf{2})$ of $SU(2) \times SU(2)$ with `$\sim$' here meaning `transforms as'. Then a Dirac spinor, $\psi$, is made from the direct sum of $\psi_L$ and $\psi_R$, $\psi \sim (\textbf{2}, \textbf{1})\oplus(\textbf{1}, \textbf{2})$. Accordingly, in four-component notation the Weyl spinors in the Weyl basis are $(\psi_L , 0)$ and $(0, \psi_R)$, and are eigenfunctions of $\gamma^5$ with eigenvalues $-1$ and $+1$, respectively.

The usual Majorana condition for a Dirac spinor has the form, $\psi = C \bar{\psi}^T$ , where $C$ is the charge-conjugation matrix. In four dimensions C is off-diagonal in the Weyl basis, since it maps the components transforming as $(\textbf{2}, \textbf{1})$ into $(\textbf{1}, \textbf{2})$. For even D, it is always possible to define a Weyl basis where $\Gamma^{D+1}$ (which consists of the product of all  matrices in D dimensions) is diagonal, therefore
\begin{equation}
\label{gammaEigenV}
    \Gamma^{D+1}\psi_{\pm}=\pm\psi_{\pm}\,.
\end{equation}
We can express $\Gamma^{D+1}$ in terms of the chirality operators in four and extra $d$ dimensions,
\begin{equation}
    \Gamma^{D+1}=\gamma^5\otimes\gamma^{d+1}\,.
\end{equation}
As a result, the eigenvalues of $\gamma^5$ and $\gamma^{d+1}$ are interrelated. It should be noted though that the choice of the eigenvalue of $\Gamma^{D+1}$ does not impose the eigenvalues on the separate $\gamma^5$ and $\gamma^{d+1}$.

Given that $\Gamma^{D+1}$ commutes with the Lorentz generators, then each of the $\psi_+$ and $\psi_-$ corresponding to its two eigenvalues, according to eq. \eqref{gammaEigenV}, transforms as an irreducible spinor of $SO(1,D-1)$. For $D$ even, the $SO(1,D-1)$ always has two independent irreducible spinors; for $D = 4n$ there are two self-conjugate spinors $\sigma_D$ and ${\sigma_d}^\prime$, while for $D = 4n + 2$, $\sigma_D$ is non-self-conjugate and $\bar{\sigma}_D$ is the other spinor. Conventionally, it is selected $\psi_-\sim \sigma_D$ and $\psi_+\sim {\sigma_D}^\prime$ or $\bar{\sigma}_D$. Then, Dirac spinors are defined as direct sum of Weyl spinors,
\begin{equation}
\psi=\psi_+\oplus\psi_-\sim
    \begin{cases}
    \sigma_D\oplus\sigma_D^{\prime} & \mathrm{for}\ D=4n \\
    \sigma_D\oplus\bar{\sigma}_D & \mathrm{for}\ D=4n+2\,.  
    \end{cases}
\end{equation}

The Majorana condition can be imposed in $D=2,3,4+8n$ dimensions and therefore the Majorana and Weyl conditions are compatible only in $D=4n+2$ dimensions. We limit ourselves here in the case that $D = 4n+2$ (or the rest see e.g. refs \cite{CHAPLINE1982461, KAPETANAKIS19924}). Then starting with Weyl–Majorana spinors in $D=4n+2$ dimensions, we are actually forcing a rep, $f_R$, of a gauge group defined in higher dimensions to be the charge conjugate of $f_L$, and we arrive in this way to a four-dimensional theory with the fermions only in the $f_L$ rep of the gauge group.

Lets discuss our example now, keeping again the Euclidean signature, with the Weyl spinor of $SO(18)$, 256 and according to the breakings and branching rules discussed earlier in the present section we have
\begin{equation}
\begin{aligned}
SO(18) &\supset SU(4)\times SO(12)\\
\textbf{256} &= (\textbf{4},\overline{\textbf{32}}) + (\overline{\textbf{4}},\textbf{32})\, .
\end{aligned}
\end{equation}
Given that the Majorana condition can also be imposed we are led to have fermions in the $(\overline{\textbf{4}},\textbf{32})$ of $SU(4) \times SO(12)$. Then we have the following branching rule of the \textbf{32} under the $SO(10) \times [U(1)]$
\begin{equation}
\label{SO12}
\begin{aligned}
SO(12) &\supset SO(10) \times [U(1)] \\
\textbf{32} &= (\overline{\textbf{16}})(1) + (\textbf{16})(-1)\, .
\end{aligned}
\end{equation}
The $[U(1)]$ is put to take into account the case that $U(1)$ exists as gauge symmetry and the case that it is broken (see the breaking of $SO(12)$ with scalar in the \textbf{77} rep in eq. \eqref{SO12}) leaving a $U(1)$ as a remaining global symmetry.

On the other hand, as noted earlier,
\begin{equation}
\begin{aligned}
SU(4) &\supset SU(2) \times SU(2) \\
\textbf{4} &= (\textbf{2},\textbf{1}) + (\textbf{1},\textbf{2})\, .
\end{aligned}
\end{equation}
Therefore, after all the breakings, we obtain:
\begin{equation}
\begin{aligned}
      SU(2) &\times SU(2) \times SO(10) \times [U(1)]\\
      \{[(\textbf{2}, \textbf{1}) &+ (\textbf{1}, \textbf{2})\} \{(\overline{\textbf{16}})(1)+(\textbf{16})(-1)\}\\
      &=\overline{\textbf{16}}_L (1)+\textbf{16}_L(-1)+\overline{\textbf{16}}_R(1)+\textbf{16}_R(-1)
\end{aligned}
\end{equation}
and since $\overline{\textbf{16}}_R(1)=\textbf{16}_L(-1)$ and $\overline{\textbf{16}}_L(1) = \textbf{16}_R(-1)$, the above expression becomes:
\begin{equation}
=2 \times \textbf{16}_L (-1)  +2  \times \textbf{16}_R (-1). 
\end{equation}
Finally, choosing to keep only the $-1$ eigenvalue of $\gamma^5$ we obtain:
\begin{equation}
    2 \times \textbf{16}_L (-1)\, .
\end{equation}

Similarly to the general discussion we presented earlier in this section concerning the Weyl condition in $D$ and 4 dimensions, namely that they are independent to each other, the same holds for the Majorana condition. If we impose the Majorana condition in higher dimensions we are still free to impose the Majorana condition once more in lower dimensions, taking into account the rule for the non-compact groups $SO(p, q)$ mentioned earlier in the present section. Therefore if we impose in addition the Weyl also the Majorana condition in higher dimensions we can still impose the same conditions in lower dimensions, respecting the known rules for each case.

Therefore, given the above analysis the gauge group describing the Internal Interactions is $C_{SO(18)} (SO(6)) = SO(10) \times U(1)_{\text{global}}$, while the type of spinors that we have is governed by the signature of $ (p-q)$ that permits the imposition of Weyl and Majorana conditions in higher and four dimensions leading to one generation of $\textbf{16}_L$ in $SO(10)$. Obviously, the other fermion generations are introduced as usual with more spinors in $SO(2, 16)$.

An additional comment is necessary concerning the case of FG. As it was explained in \cite{roumelioti2407}, when attempting to unify FG with internal interactions, along the lines of Unification of CG with $SO(10)$ \cite{Roumelioti:2024lvn}, the difficulties that in principle one is facing are that fermions should (a) be chiral in order to have a chance to survive in low energies and not receive masses as the Planck scale, (b) appear in a matrix representation, since the constructed FG is a matrix model. Then, given that the Majorana condition can be imposed, a solution satisfying the conditions (a) and (b) above was suggested in \cite{roumelioti2407}: We choose to start with $SO(6) \times SO(12)$ as the initial gauge theory with fermions in the $(\textbf{4}, \overline{\textbf{32}})$ representation satisfying in this way the criteria to obtain chiral fermions in tensorial representation of a fuzzy space. Another important point is that using the gauge-theoretic formulation of gravity to construct the FG model, one is led to gauge the $SO(6) \times U(1) \sim SO(2, 4) \times U(1)$. Therefore, from this point of view, there only exists a small difference in comparison to the CG case.

\section{From SO(2,16) to the Standard Model}
\label{sec7}

In this section four distinct models that begin from the $SO(2,16)\sim SO(18)$ gauge group and result in the SM are examined, as well as their observation potential in experiments that search for gravitational wave signals and  proton decay.

\subsection{Field content and estimation of scales of spontaneous symmetry breakings}\label{content-scales}

First, we determine (following \cite{Patellis:2024znm}) the full field content of the initial $SO(18)$ gauge theory, from which we ultimately obtain EG and $SO(10)\times [U(1)]_{global}$.  We follow the breaking directions and necessary field content employed in \cite{Djouadi:2022gws}, in order to finally result in the SM. More specifically, $SO(10)$ breaks into an intermediate gauge group, which in turn breaks into the SM group. The intermediate gauge groups are the Pati-Salam, $SU(4)_C\times SU(2)_L\times SU(2)_R$, with or~without a discrete left-right symmetry,~$\mathcal{D}$ and the minimal~left-right group, $SU(3)_C\times SU(2)_L\times SU(2)_R\times U(1)_{B-L}$, with or without a discrete left-right symmetry. From now on,  we  denote them 422, 422D, 3221 and 3221D, respectively. Naturally, we have four different field contents at the $SO(18)$ level, one for each specific lower-energy model.

As explained in detail in the previous section, $SO(18)$ breaks into $SO(6)\times SO(12)$ by the $(\textbf{1},\textbf{1})$ of a scalar $\textbf{170}$ rep, while we choose scalars in the $\textbf{15}$ rep of~$SO(6)$ to break the CG group, which can be drawn from the $SO(18)$ rep $\textbf{153}$:
\begin{equation}\label{153}
\textbf{153}   = (\textbf{15},\textbf{1}) + (\textbf{6},\textbf{12}) + (\textbf{1},\textbf{66})~,
\end{equation}
The $SO(12)$ gauge group is broken (spontaneously) by scalars in the $\textbf{77}$ rep, which can result from a $\textbf{170}$ rep of the parent group. 
Therefore, in $SO(6)\times SO(12)$ notation, the scalars responsible for the breakings of the two groups belong to $(\textbf{15},\textbf{1})$ and $(\textbf{1},\textbf{77})$. In order to have fermions in the $\textbf{16}$ rep of $SO(10)$, we need three copies of a $\textbf{256}$ rep of $SO(18)$ (which will result in the $\textbf{16}$ through $(\overline{\textbf{4}},\textbf{32})$ in $SO(6)\times SO(12)$ notation). The $SO(10)$ GUT is broken by a scalar  in the  $\textbf{210}$ rep into the 422 and the 3221D gauge groups, by a scalar in the $\textbf{54}$ rep into the 422D gauge group and by a scalar in the  $\textbf{45}$ rep  into the 3221 gauge group.

All intermediate gauge groups break spontaneously into the SM with scalars in a $\overline{\textbf{126}}$ rep and the electroweak Higgs boson is accommodated in a $\textbf{10}$ rep\footnote{In order for the Higgs boson to be in a $\textbf{10}$ rep instead of a $\textbf{120}$ and to~avoid an additional Yukawa term, a $U(1)$ Peccei-Quinn symmetry must be taken into account \cite{Djouadi:2022gws}. This could be identified with the global $U(1)$ of $SO(10)\times [U(1)]_{global}$, which also breaks at the unification scale. For the purposes of the present study this global $U(1)$ will be ignored from now on.} (stil  in $SO(10)$ language).
From now on, the scale at which the $SO(10)$ gauge group breaks will be called GUT scale, $M_{GUT}$, As the gauge couplings unify at that scale, while the scale at which the 422(D)/3221(D) groups break will be called \textit{intermediate scale}, $M_I$. The respective breakings are given below:

\footnotesize
\begin{align}
\text{422}: & \quad \text{SO(10)}|_{M_{GUT} } \xrightarrow{\langle \mathbf{210_H} \rangle} \, \, \, SU(4)_C\times SU(2)_R\times SU(2)_R|_{M_{I} }\xrightarrow{\langle \mathbf{\overline{126}_H}\rangle} \text{SM}\, ; \label{breakingchain1} \\
\text{422D}: & \quad \text{SO(10)}|_{M_{GUT} } \xrightarrow{\langle \mathbf{54_H} \rangle} ~~SU(4)_C\times SU(2)_R\times SU(2)_R \times {\cal D}|_{M_{I}}\xrightarrow{\langle \mathbf{\overline{126}_H}\rangle
} \text{SM}  \, ; \label{breakingchain2} \\
\text{3221}: & \quad \text{SO(10)}|_{M_{GUT}} \xrightarrow{\langle \mathbf{45_H} \rangle} \, \, \, SU(3)_C\times SU(2)_L\times SU(2)_R\times U(1)_{B-L}|_{M_{I}}\xrightarrow{\langle \mathbf{\overline{126}_H} \rangle} \text{SM}\, ; \label{breakingchain3}
\\
\text{3221D}: & \quad \text{SO(10)}|_{M_{GUT}} \xrightarrow{\langle \mathbf{210_H} \rangle} SU(3)_C\times SU(2)_L\times SU(2)_R\times U(1)_{B-L}\times {\cal D}|_{M_{I}}\xrightarrow{\langle \mathbf{\overline{126}_H} \rangle} \text{SM} \, .
\label{breakingchain4}
\end{align}
\normalsize
After examining the following branching rules: 
\begin{align} 
     SO(12) &\supset SO(10) \times [U(1)]_{global} \nonumber\\
   \textbf{12}   &= (\textbf{1})(2) + (\textbf{1})(-2) + ( \textbf{10})(0) \label{12}\\
   \textbf{66}   &= (\textbf{1})(0) + (\textbf{10})(2) +(\textbf{10})(-2) + ( \textbf{45})(0) \label{66}\\
   \textbf{77}   &= (\textbf{1})(4) +(\textbf{1})(0) +(\textbf{1})(-4) + (\textbf{10})(2) +(\textbf{10})(-2) + ( \textbf{54})(0) \label{77}\\
   \textbf{495}   &= (\textbf{45})(0)  + (\textbf{120})(2) +(\textbf{120})(-2) + ( \textbf{210})(0) \label{495}\\
   \textbf{792}   &= (\textbf{120})(0) +(\textbf{126})(0) +(\overline{\textbf{126}})(0) + (\textbf{210})(2) +(\textbf{210})(-2)\label{792}~,
\end{align}
we accommodate~the Higgs $\textbf{10}$ rep into the  $\textbf{12}$ rep of $SO(12)$, while the $\overline{\textbf{126}}$ scalars that break the intermediate gauge group are accommodated into $\textbf{792}$. Regarding the intermediate breakings, $\textbf{45}$  comes from $\textbf{66}$, $\textbf{54}$ comes from $\textbf{77}$ and $\textbf{210}$ comes from $\textbf{792}$. Considering the $SO(18)$ branching rules:
\begin{align} 
     SO(18) \supset& SO(6) \times SO(12) \nonumber\\
   \textbf{18}   =& ( \textbf{6}, \textbf{1}) + (\textbf{1}, \textbf{12})\label{18}\\
   \textbf{3060}   =& (\textbf{15},\textbf{1}) + (\textbf{10},\textbf{12}) + (\overline{\textbf{10}},\textbf{12}) + (\textbf{15},\textbf{66}) + (\textbf{6},\textbf{220}) + (\textbf{1},\textbf{495})\label{3060}\\
   \textbf{8568}   =& (\textbf{6},\textbf{1}) + (\textbf{15},\textbf{12})+ (\textbf{10},\textbf{66}) + (\overline{\textbf{10}},\textbf{66}) + (\textbf{15},\textbf{220}) + (\textbf{6},\textbf{495}) + \nonumber\\
   &+   (\textbf{1},\textbf{792})\label{8568}~,
\end{align}
and the branching rules of (\ref{so18}) and (\ref{153}), the $\textbf{12}$ rep of $SO(12)$ comes from $\textbf{18}$ of $SO(18)$, $\textbf{792}$ comes from $\textbf{8568}$, $\textbf{66}$~from $\textbf{153}$, $\textbf{495}$ from $\textbf{3060}$ and finally $\textbf{77}$ from $\textbf{170}$.   The full field content under the reps of each gauge group is  given in Table \ref{content}.
\begin{center}
\begin{table}
\begin{center}
\renewcommand{\arraystretch}{2}
\begin{tabular}{|r|r|r|r|}
\hline
    $SO(10)$ & $SO(6)\times SO(12)$ & $SO(18)$  & Type  \& Role  \\\hline
 $\textbf{16}$ & $(\overline{\textbf{4}},\textbf{32})$ & $\textbf{256}$  & fermion, 3x generations   \\\hline
 - & $(\textbf{15},\textbf{1})$ &  $\textbf{153}$ & scalar, breaks $SO(6)$   \\\hline
 - & $(\textbf{1},\textbf{77})$ & $\textbf{170}$ & scalar, breaks $SO(12)$   \\\hline
 $\textbf{18}$ & $(\textbf{1},\textbf{12})$ & $\textbf{18}$$\textbf{18}$ & scalar, breaks SM   \\\hline
 $\overline{\textbf{126}}$ & $(\textbf{1},\textbf{792})$ & $\textbf{8568}$ & scalar, breaks the intermediate  groups into SM  \\\hline
 $\textbf{45}$ & $(\textbf{1},\textbf{66})$ & $\textbf{153}$ & scalar, breaks $SO(10)$ into 3221   \\\hline
 $\textbf{210}$ & $(\textbf{1},\textbf{495})$ & $\textbf{3060}$ & scalar, breaks $SO(10)$ into 422 \& 3221D   \\\hline
 $\textbf{54}$ & $(\textbf{1},\textbf{77})$ & $\textbf{170}$ & scalar, breaks $SO(10)$ into 422D   \\\hline
\end{tabular}
\caption{The full field content the respective rep under each group. 
}
\label{content}
\end{center}
\end{table}
\end{center}
We proceed with an estimation of the scales where the above-mentioned breakings~occur, by running the gauge couplings at each energy regime, using 1-loop renormalization group equations (RGEs).

\begin{figure}
\centering
\includegraphics[width=.4\textwidth]{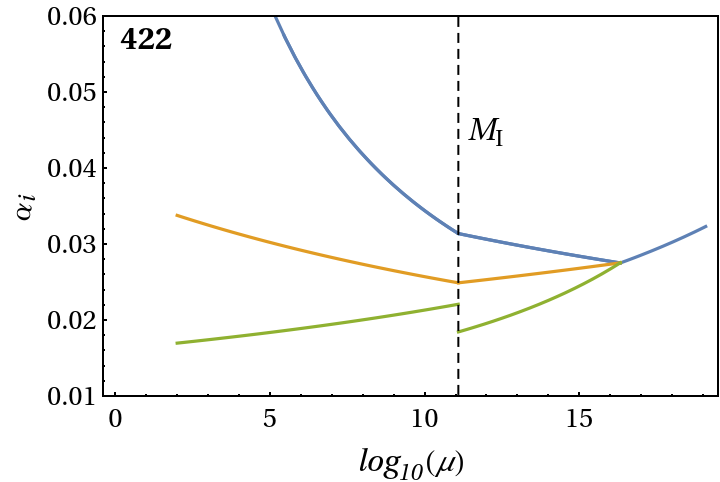}
\includegraphics[width=.4\textwidth]{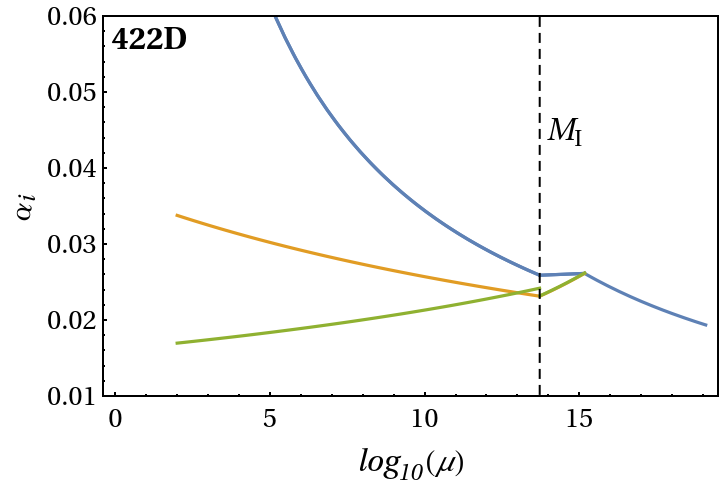}
\includegraphics[width=.4\textwidth]{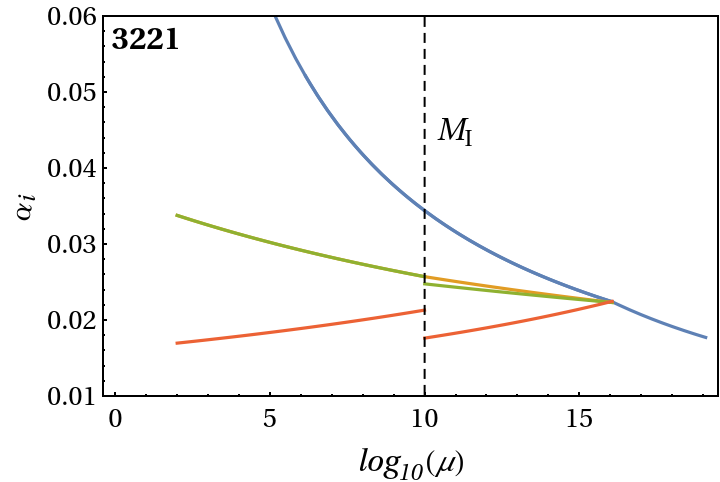}
\includegraphics[width=.4\textwidth]{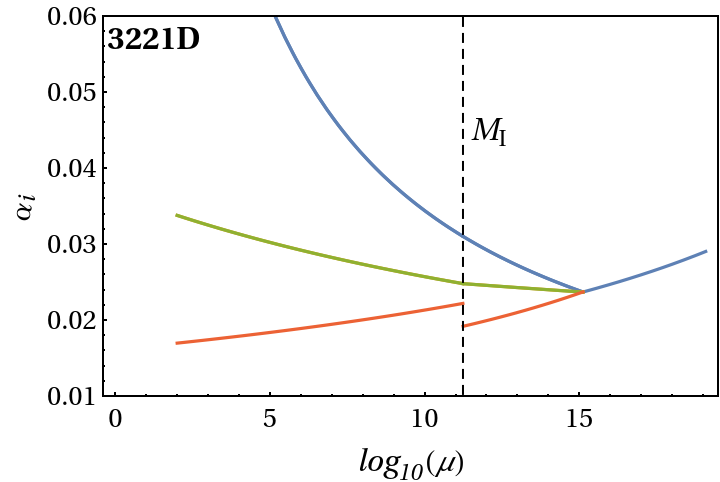}
\caption{The RG evolution of!gauge couplings from the EW scale up to almost the Planck scale is given for the four scenaria. Top left: 422;~Top right: 422D; Bottom~left: 3221; Bottom~right: 3221D.}
\label{low-ene}
\end{figure}

We start from the SM and specifically the $M_Z$, where the values of the three gauge couplings are known from experiments  \cite{ParticleDataGroup:2022pth}. Then,~using the 1-loop gauge $\beta$-functions of the SM energy regime and of each of the intermediate groups (calculated in \cite{Patellis:2024znm}), we can find the intermediate scale $M_I$ and the GUT scale that allow gauge unification, but also the value of the unified~gauge coupling at that scale, $g_{10}(M_{GUT})$.  These results can be found in Table \ref{low-energy}. The matching conditions for the  breaking of 422 to the SM at $M_I$ are:
\begin{align*}
\alpha_4^{422}(M_I)=&\alpha_3^{SM}(M_I)\\
\alpha_{2L}^{422}(M_I)=&\alpha_2^{SM}(M_I)\\
\frac{1}{\alpha_{2R}^{422}(M_I)}=&-\frac{2}{3}\frac{1}{\alpha_{3}^{SM}(M_I)}+\frac{5}{3}\frac{1}{\alpha_{1}^{SM}(M_I)}~,
\end{align*}
while the~matching conditions for the breaking of 3221 are:
\begin{align*}
\alpha_3^{3221}(M_I)=&\alpha_3^{SM}(M_I)\\
\alpha_{2L}^{3221}(M_I)=&\alpha_{2R}^{3221}(M_I)=\alpha_2^{SM}(M_I)\\
\frac{1}{\alpha_{1}^{3221}(M_I)}=&\frac{5}{2}\frac{1}{\alpha_{1}^{SM}(M_I)}-\frac{3}{2}\frac{1}{\alpha_{2}^{SM}(M_I)}~.
\end{align*}
Their RG evolution of each of the four cases is given in Figure \ref{low-ene}.
\begin{center}
\begin{table}
\begin{center}
\renewcommand{\arraystretch}{2} 
\begin{tabular}{|l|r|r|r|}
\hline
       & $M_I$ (GeV) & $M_{GUT}$ (GeV) & $g_{10}^{(1)}(M_{GUT})$  \\\hline
 422   & $1.2\times10^{11}$ & $2.1\times10^{16}$  & $0.587$   \\\hline
 422D  & $5.2\times10^{13}$ & $1.5\times10^{15}$  & $0.572$   \\\hline
 3221  & $1.0\times10^{10}$ & $1.1\times10^{16}$  & $0.531$   \\\hline
 3221D & $1.7\times10^{11}$ & $1.4\times10^{15}$  & $0.546$   \\\hline
\end{tabular}
\caption{1-loop results for the intermediate scale, the unification scale and the unified gauge coupling at $M_{GUT}$. 
}
\label{low-energy}
\end{center}
\end{table}
\end{center}
It should be noted that the~breaking of the CG to~EG gives a negative~contribution to the cosmological constant~and, if this was the only contribution,~the space would be AdS. Thankfully, we have positive contributions from the  breakings of $SO(18)$ and $SO(12)$. If we choose either of these spontaneous breakings to happen at the same scale as the CG breaking, then the  contributions can be fine-tuned to give a zero or slightly positive value for the cosmological constant in agreement with  experimental observations.

\begin{figure}
\centering
\includegraphics[width=.8\textwidth]{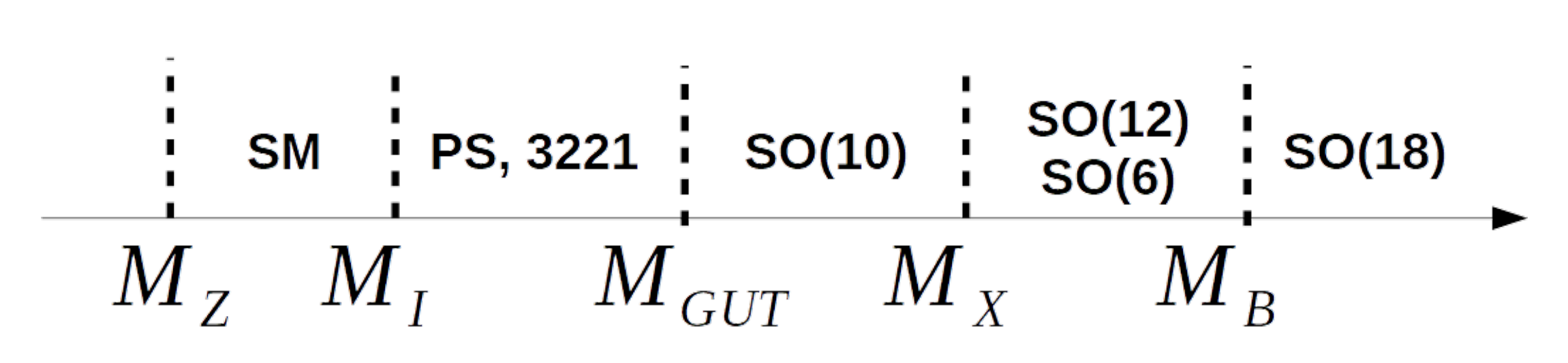}
\caption{The hierarchy among the  symmetry breaking scales of the models and the gauge groups that hold on each energy regime. For scenario \textit{A} we have $M_X=M_B$, for scenario \textit{B} we have $M_B<M_{Pl}$, while for scenario \textit{C} we have $M_B=M_{Pl}$.}
\label{scales}
\end{figure}

We focus on three different scenaria regarding the breakings beyond the GUT scale. In scenario \textit{A} the $SO(18)$  group breaks into~$SO(6)\times SO(12)$ and they in turn break into EG and $SO(10)$  all at the same scale,  $M_X$. As such,  the contribution to the cosmological constant from the breaking of $SO(18)$ cancels the negative one originating from the CG breaking. On the other hand, in the scenaria  \textit{B} and \textit{C}, $SO(18)$  breaks~into $SO(6)\times SO(12)$ at a scale $M_B$, while both~$SO(6)$ and $SO(12)$ will~break at another scale, $M_X$, lower than $M_B$ but higher than $M_{GUT}$. In both \textit{B} and \textit{C} it is the contribution to the cosmological constant from the breaking of $SO(12)$ cancels the negative one.
The above are visualised  in Figure \ref{scales}.

The RG evolution of the couplings below the GUT scale is clear and we calculate the RGEs as usual. On the contrary, regarding the running of gauge theories~based on non-compact~groups, the situation is certainly not that clear. There exist very serious calculations of the $\beta$-functions~of the various~terms of Stelle's $R^2$ gravity, which was~proven to be renormalizable \cite{Stelle:1976gc,Stelle:1977ry}.  However, all calculations are done in~Euclidean space~\cite{Fradkin:1981iu,Avramidi:1985ki,Codello:2006in,Niedermaier:2009zz,Niedermaier:2010zz,Ohta:2013uca}. Therefore,~strictly speaking, the calculation~of the $\beta$-function of a gauge theory~based on a non-compact~group has not been~done. We speculate, though, that at~least at~one-loop level, the $\beta$-functions of~gauge theories based on such groups could be well~approximated by the respective ones of their compact counterparts. This finds support from~suggestions of Donoghue \cite{Donoghue:2016xnh,Donoghue:2016vck,Donoghue:2017fvm}, which we adapt when calculating $\beta$-functions (see \cite{Patellis:2024znm})).\\

\noindent \textbf{Scenario \textit{A}}:  the $SO(10)$ gauge coupling runs until the $M_X$ scale, where it has to match the value of both the $SO(6)$ and~$SO(12)$ gauge couplings:
\begin{equation}
\alpha_{10}^{(1)}(M_X)=\alpha_{CG}^{(1)}(M_X)~. \label{mc}
\end{equation}
Substituting the above relation into (\ref{so24finalaction}) and considering its last term, we can compare this term with the $SO(12)$ contributions to the cosmological~constant. Thus, we have an estimate of the scale (its precise~value will depend~on various~parameters):
\begin{equation}
    M_X \sim  10^{18}~\text{GeV}~. \label{MxA}
\end{equation}
Unfortunately, if we try to run the $SO(18)$ gauge coupling up to $M_{Pl}$, it becomes clear that its steep $\beta$-function pushes it rapidly in the non-perturbative~regime and it has a Landau pole before it reaches the Planck scale. A drastic change of the field content and/or additional new physics phenomena below $M_{Pl}$ could in principle ameliorate the situation.\\

\noindent \textbf{Scenaria \textit{B} \& \textit{C}}:  in the former we break $SO(18)$ below the Planck scale, $M_B<M_{Pl}$, while in the latter $SO(18)$ breaks at the Planck scale, $M_B=M_{Pl}$. Consequently, in both scenaria, the $SO(6)\times SO(12)$~gauge group runs until $M_X$, below which only $SO(10)$  and EG are left (and the global $U(1)$ that ignore throughout the study). We remind that $SO(6)$ and $SO(12)$ should always break at the same scale, in order to fine tune the cosmological constant.

We cannot use a matching condition like  \refeq{mc} in either case, as we~now have $\alpha_{10}^{(1)}(M_X)=\alpha_{12}^{(1)}(M_X)$. Therefore, employing both the $SO(10)$ and $SO(12)$ gauge $\beta$-functions  \textit{and} the approximative gauge $\beta$-function of $SO(6)$, we can make a rough estimate of the scale, which is once more  $M_X\sim10^{18}$GeV. 

\begin{figure}
\centering
\includegraphics[width=.4\textwidth]{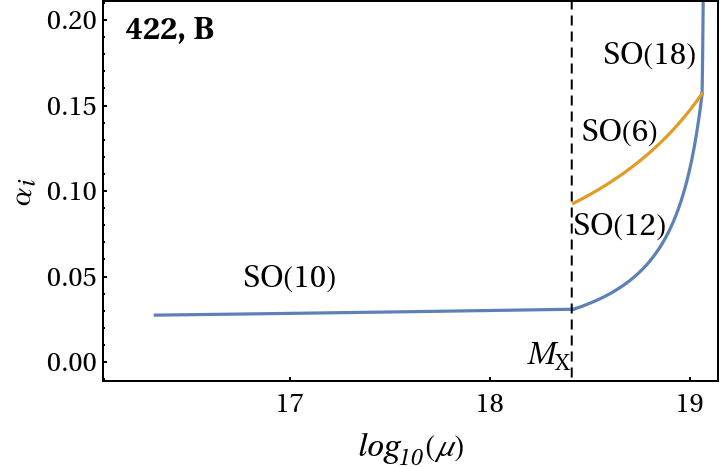}
\includegraphics[width=.4\textwidth]{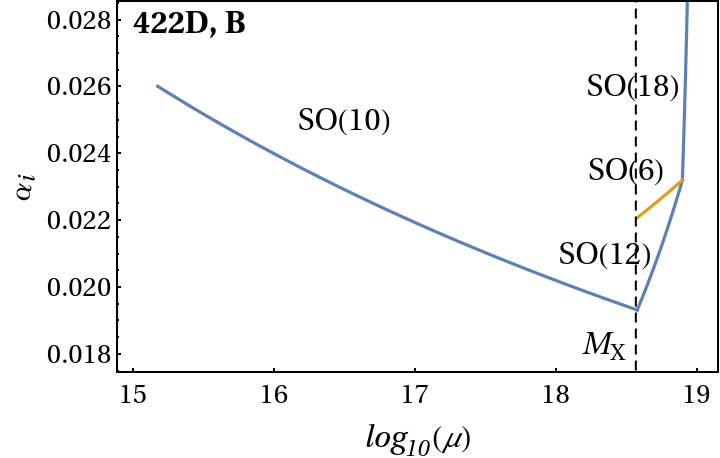}
\includegraphics[width=.4\textwidth]{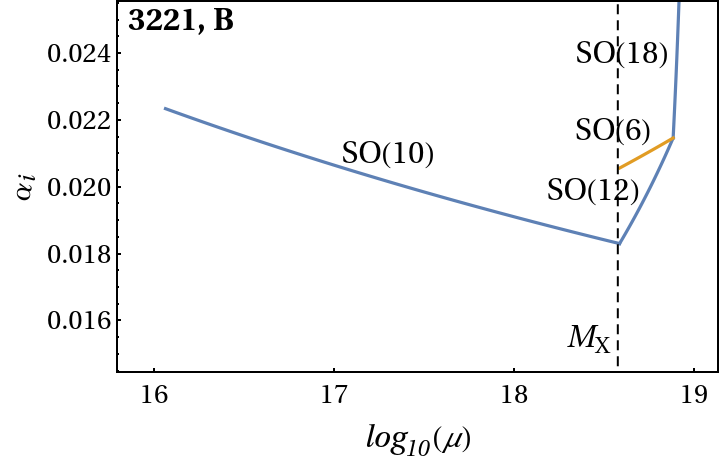}
\includegraphics[width=.4\textwidth]{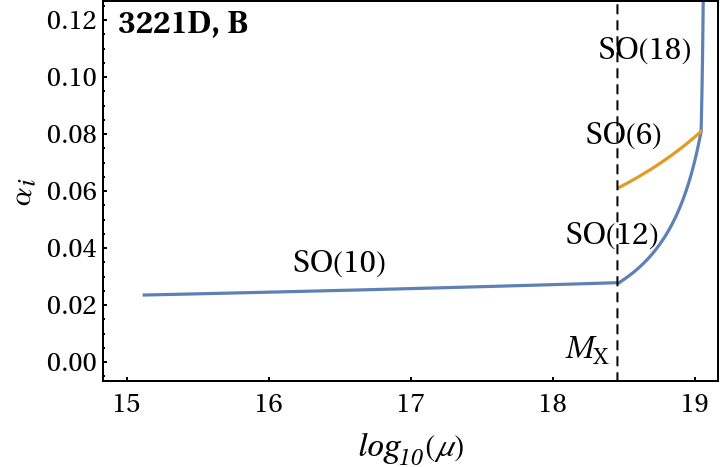}
\caption{The gauge RG evolution for scenario \textit{B}. Top left: 422; Top right: 422D; Bottom left: 3221; Bottom right: 3221D.}
\label{b}
\end{figure}

\begin{figure}
\centering
\includegraphics[width=.4\textwidth]{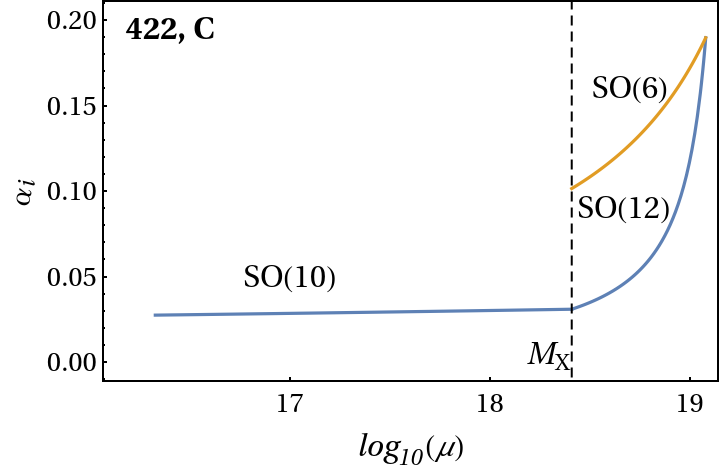}
\includegraphics[width=.4\textwidth]{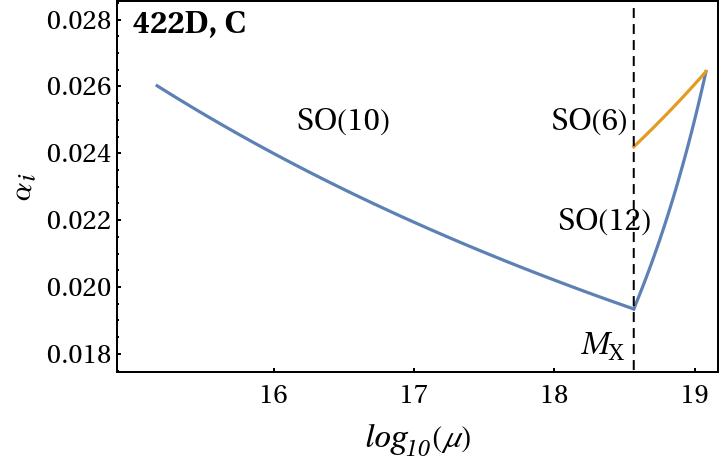}
\includegraphics[width=.4\textwidth]{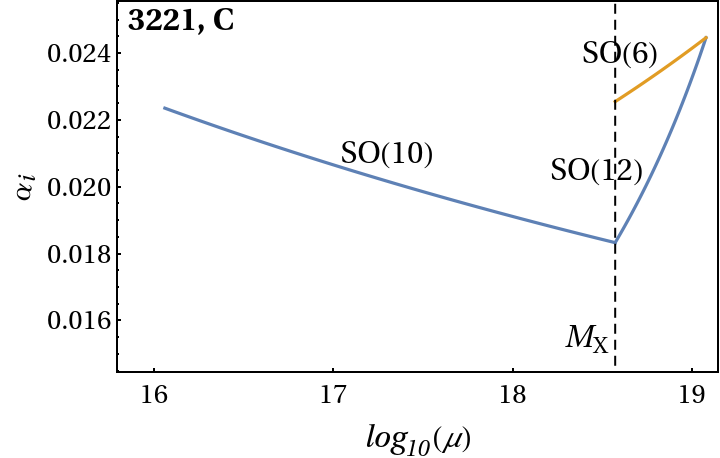}
\includegraphics[width=.4\textwidth]{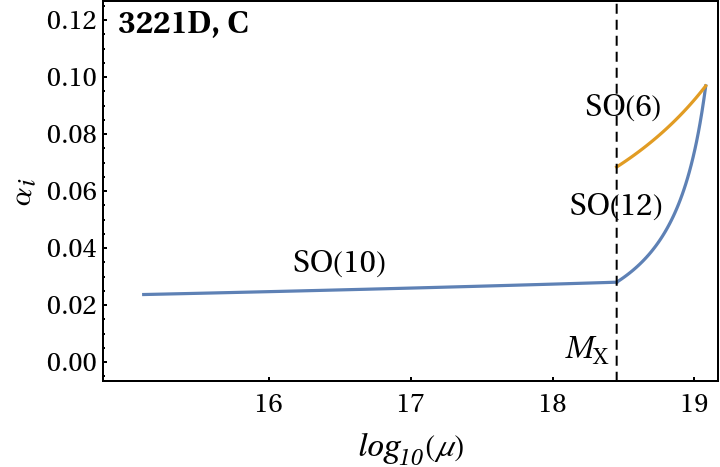}
\caption{The gauge RG evolution for scenario \textit{C}. Top left: 422; Top right: 422D; Bottom left: 3221; Bottom right: 3221D.}
\label{c}
\end{figure}

Above $M_X$ the gauge coupling of $SO(12)$ runs up to $M_B$ staying within the perturbative regime. This is due to our choice of reps for some of the scalars. More specifically, we chose them in such a way that the scalars are always singlets under the CG gauge group, thus avoiding any multiplicities in the calculation of the gauge $\beta$-function of $SO(12)$. In scenario  \textit{B} the gauge coupling of $SO(18)$  can run up to $M_{Pl}$ staying in the perturbative regime, although it features a very steep RG evolution. The above can be found in Figures \ref{b} and \ref{c}.

A comment on the FG case is in order. The attempt to unify FG with internal interactions (along the lines of \cite{Roumelioti:2024jib} and \cite{Roumelioti:2024lvn})
comes with several difficulties. In particular, fermions should be chiral in order to not acquire Planck scale masses and also they should appear in a matrix rep, since FG is a matrix model.
Since one can impose the Majorana condition, a solution to the above is that we start with the $SO(6) \times SO(12)$ group as the initial gauge theory and the fermions are accommodated in the $(\overline{\textbf{4}},\textbf{32})$ rep.
Furthermore, following the gauge-theoretic formulation of gravity in order to construct a FG, we gauge $SO(6) \times U(1) \sim SO(2,4) \times U(1)$. As such, it is very similar to the  CG case, since the abelian symmetry is irrelevant to the above study and we could identify the FG model to scenario \textit{C}, with the (notable but irrelevant) difference that there is no $SO(18)$ gauge group above $M_{Pl}$.

\subsection{Cosmic Strings from intermediate scale spontaneous symmetry breaking and proton decay constraints}\label{cosmic-strings}

Historically, the first important test for every GUT is proton decay. While it has not yet been observed, the proton lifetime has been at the center of many experimental searches \cite{Super-Kamiokande:2013rwg, Super-Kamiokande:2014otb, Super-Kamiokande:2016exg, Heeck:2019kgr}, which severely constrain it and, in turn, the unification scale. Most of the breaking directions of the $SO(10)$ GUT are rigorously tested by Super-Kamiokande (Super-K), while future experiments -like Hyper-Kamiokande (Hyper-K) \cite{Hyper-Kamiokande:2018ofw}, DUNE \cite{DUNE:2016hlj} and JUNO \cite{JUNO:2015zny}- will improve  sensitivity by even one order of magnitude. Given the predictions of many GUTs (lifetime below $10^{36}$ years), the above-mentioned experiments could be getting very close to the proton decay discovery and, consequently, baryon  number violation. This would be a significant milestone for theoretical particle physics and it would also exclude many attempts to grand unification.

However, proton decay is not the only way to probe GUTs in current and near future experiments. When the GUT gauge symmetry or any of the intermediate gauge structures break spontaneously towards the SM, they produce topological defects. Some of them, namely domain walls and monopoles, dominate the energy density of the Universe, and are therefore considered problematic. This problem can fortunately go away with the assumption that inflation takes place after their production, as it suppresses their density strongly. A third type of defect, cosmic strings, which are formed from the breaking of an abelian $U(1)$~subgroup (although in some cases larger subgroups can result in cosmic strings), do not present such a problem, as a cosmic string network has a scaling solution and thus does~not overclose the~Universe. It can survive, though, and generate~a source of~gravitational radiation~\cite{Vilenkin:1984ib, Caldwell:1991jj, Hindmarsh:1994re}. 

Recently, gravitational waves (GWs) originating from cosmic strings have been highlighted as a way to probe high energy  models (i.e. GUTS) \cite{Damour:2001bk, Damour:2004kw, Dror:2019syi, Buchmuller:2019gfy, Chakrabortty:2020otp, King:2020hyd}. Assuming that we have inflation before the formation of cosmic strings, when they intersect and form loops their network acts as  a GW source. When transitioning between different states, they emit~strong beams of GW of high frequency. Additionally, the loops oscillate,~shrink~and emit~energy in a gravitational manner. This radiation is a stochastic gravitational~wave background (SGWB). An approach to compute SGWB can be found in \cite{Cui:2018rwi}, where Nambu-Goto strings are assumed that decay predominantly via gravitational~radiation. The decay of cosmic strings of the type discussed above releases energy that can be transferred to gravitational~radiation. 

Regarding Nambu-Goto strings, the dominant contribution to GW signal comes from large loops and thus we focus on them. After the formation of the strings, the loops emit gravitational radiation~at a constant~rate:
\begin{equation}
\frac{dE}{dt}=-\Gamma G\mu^2~,
\end{equation}
where $G$ is the Newton constant, $\mu$ is the string tension and $\Gamma\sim50$ \cite{Blanco-Pillado:2017oxo}. The value of $G\mu$ can be explored and constrained by current and near future GW ~detectors. in the Hz regime, LIGO~(O3) \cite{LIGOScientific:2019vic} excludes the formation of cosmic strings at $G\mu \sim10^{-8}$ in the high-frequency range of $10-100$ Hz. In the nHz range, EPTA~\cite{EPTA:2015qep} and NANOGrav~\cite{NANOGRAV:2018hou} set an upper bound of $G\mu$ at $6 \times 10^{-11}$. However, the strongest constraint comes from
the PPTA collaboration \cite{Blanco-Pillado:2017rnf}, setting an upper bound  at $1.5 \times 10 ^{-11}$.

In our study in Subsection \ref{content-scales} we have chosen four breaking chains that lead from the $SO(10)$  GUT to the SM along the lines of \cite{Djouadi:2022gws}, denoted 422, 422D, 3221 and 3221D respectively, and we have calculated the intermediate scale $M_I$ and the unification scale $M_{GUT}$ for each case, using 1-loop $\beta$-functions for minimal field content (the results are given in Table \ref{low-energy}). In \cite{King:2021gmj} one can find a 2-loop calculation of intermediate and unification scales for all the breaking chains of $SO(10)$ with one intermediate breaking, including the four cases we use in our study. We follow their comprehensive analysis, which compares their numerical results  regarding the respective scales to the experimental bounds on proton decay/lifetime of Super-K and Hyper-K. It turns out that neither 422D  nor 3221D satisfy the bounds set by Super-K, while 422 and 3221 are just above the lower bounds for $M_{GUT}$. However, if during the 10-year exposure~time of the future Hyper-K~experiment proton decay~is not~observed, 3221 will be excluded as  well. The above leave 422 as the candidate with the highest  survivability against the stringent constraints of the Kamiokande experiments.

Regarding  the production of topological defects from the various breakings of the four cases (see \cite{King:2021gmj} for a complete categorization), the breaking of the $SO(10)$ gauge group leads to monopoles in all cases and additional cosmic strings in 422D and 3221D. This means that inflation should happen after this breaking to prevent monopoles from dominating the Universe's  energy density and, even in the two cases we have cosmic strings, they will be washed out from inflation and their gravitational signal is rendered undetectable.\footnote{This means that any gravitational signal originating from the breakings at $M_X$ and $M_B$ will also be completely diluted by inflation.} If we consider the intermediate breakings of each case, we have the production of monopoles from the breaking of the 422 gauge group, monopoles and domain walls from the 422D breaking, cosmic strings from 3221 and cosmic strings and domain walls from 3221D. With the same reasoning as above, inflation should happen after the breakings of 422, 422D and 3221D, and thus cannot be probed  through GW background. However, in the 3221 case, inflation can strategically be placed between the breaking of the GUT and the 3221 gauge group, and thus strings can in principle be observed through SGWB.

Following the analysis of \cite{King:2021gmj}, the tension of cosmic strings generated from the breaking of 3221 to the SM is given approximately by
\begin{equation}
G\mu\simeq
    \frac{1}{2(\alpha_{2R}(M_I)+\alpha_{1}(M_I))}\frac{M_I^2}{M_{Pl}^2}, 
\end{equation}
where $\alpha_{2R}(M_I)$ and $\alpha_{1}(M_I)$ are the gauge~coefficients of $SU(2)_R$ and $U(1)_{B-L}$ at $M_I$, respectively. Substituting the results of the 2-loop analysis, the tension is predicted at $G\mu \simeq 2.0  \times 10^{-17}$ and is compatible with the bounds set by Super-K. Again, should Huper-K not observe proton decay, this channel is excluded and any Gw signal should not be associated with it.

From the above we summarize that the channels 422D and 3221D are excluded from Super-K data, while 422 survives both Super-K and Hyper-K, but has no GW signal and thus cannot be probed through observation of SGWB. 3221 satisfies the Super-K proton lifetime bounds and produces  cosmic  strings that emit potentially detectable GW signal. However, it could be excluded by the non-observation of proton decay in Hyper-K. Even if observed, the signal has no memory of the discussion above the unification scale, so CG cannot be probed this way.

\section{Conclusions}
\label{sec8}
In the present paper, we have presented a quite complete scenario of the possible unification of gravity with internal interactions. It is based mostly on the suggestion of \cite{Roumelioti:2024lvn}, namely that such a unification can be achieved by gauging an enlarged tangent Lorentz group. The latter possibility results as a positive option of the observation of \cite{Weinberg:1984ke}, that the dimension of the tangent space is not necessarily equal to the dimension of the corresponding curved manifold. The scenario described here is also based on the very attractive fact that gravitational theories can be described by gauge theories as the SM of Particle Physics pointing in further examination of a possible common origin of gravitational and internal interactions. The description of gravity as gauge theory raises the question of the equivalence among diffeomorphism and gauge invariance. Indeed, an infinitesimal correspondence can be guaranteed if certain conditions are imposed. This issue has been discussed rather widely in the present paper.

The gravity theories that have been discussed here are the CG and the FG, both based on gauging of the conformal group $SO(2,4)$ (FG requires the gauging of an additional $U(1)$). Very interesting findings are that CG can be spontaneously broken to either EG, or WG, which, in a similar way can eventually be broken to EG.

Furthermore, we examine in detail the unification of both CG and FG with internal interactions in four dimensions, using the higher-dimensional tangent group $SO(2,16)$, which is eventually led, after various SSBs, to EG and the $SO(10)$ GUT. Inclusion of fermions and suitable application of the Weyl and Majorana conditions result in a fully unified scheme. This is further studied by a 1-loop analysis at low energies, employing four channels of breaking the $SO(10)$ down to the SM of Elementary Particles. Estimates of all breaking scales from the Planck scale down to the EW scale are calculated. 

Finally, the observation potential of each breaking channel in experiments that search for gravitational wave signals and proton decay is examined, following past analyses. Two of the channels are excluded by proton lifetime bounds. One of the two surviving ones produces cosmic strings that emit a detectable stochastic gravitational wave background signal and thus has the best chances of observation.

\section*{Acknowledgements}
It is a pleasure to thank Costas Bachas, Thanassis Chatzistavrakidis, Alex Kehagias, Tom Kephart, Spyros Konitopoulos, Dieter Lust, George Manolakos, Pantelis Manousselis, Carmelo Martin, Tomas Ortin, Roberto Percacci, Manos Saridakis and Nicholas Tracas, for our discussions on various stages of development of the theories presented in the current work.

DR would like to thank NTUA for a fellowship for doctoral studies. GZ would like to thank the Arnold Sommerfeld Centre - LMU Munich for their hospitality and support, the University of Hamburg and DESY for their hospitality, and the CLUSTER of Excellence "Quantum Universe" for support. GP would like to thank the Institute of Physics of U.N.A.Mexico for their warm hospitality.

\printbibliography

\end{document}